\documentclass[10pt,journal,compsoc]{IEEEtran}
\usepackage{amsmath,amsfonts}
\usepackage{algorithmic}
\usepackage{algorithm}
\usepackage{array}
\usepackage[caption=false,font=normalsize,labelfont=sf,textfont=sf]{subfig}
\usepackage{textcomp}
\usepackage{stfloats}
\usepackage{url}
\usepackage{verbatim}
\usepackage{graphicx}
\usepackage{cite}
\graphicspath{{figures/}}
\DeclareGraphicsExtensions{.pdf,.jpg,.jpeg,.png}
\usepackage{hyperref}
\usepackage{xcolor}
\usepackage[capitalise]{cleveref}
\usepackage{amsfonts}
\usepackage{amsmath}
\usepackage{tabularx}
\usepackage{multirow}
\usepackage{hhline}
\usepackage{tikz}
\usetikzlibrary{positioning,fit,calc}
\usepackage[shortcuts]{extdash}

\makeatletter
\newcommand{\gettikzxy}[3]{%
  \tikz@scan@one@point\pgfutil@firstofone#1\relax
  \edef#2{\the\pgf@x}%
  \edef#3{\the\pgf@y}%
}
\makeatother

\begin{document}

\title{Adaptive Sampling of 3D Spatial Correlations \\for Focus+Context Visualization
}

\author{Christoph~Neuhauser, 
        Josef~Stumpfegger,
        and~R\"udiger~Westermann
\thanks{The authors are with Technical University of Munich (TUM).\protect\\E-mail: \{christoph.neuhauser\,$|$\,westermann\}@tum.de, ga87tux@mytum.de.}
}


\IEEEpubid{0000--0000/00\$00.00~\copyright~2023 IEEE}

\maketitle

\begin{figure*}[ht]
\centering
\includegraphics[width=0.995\linewidth]{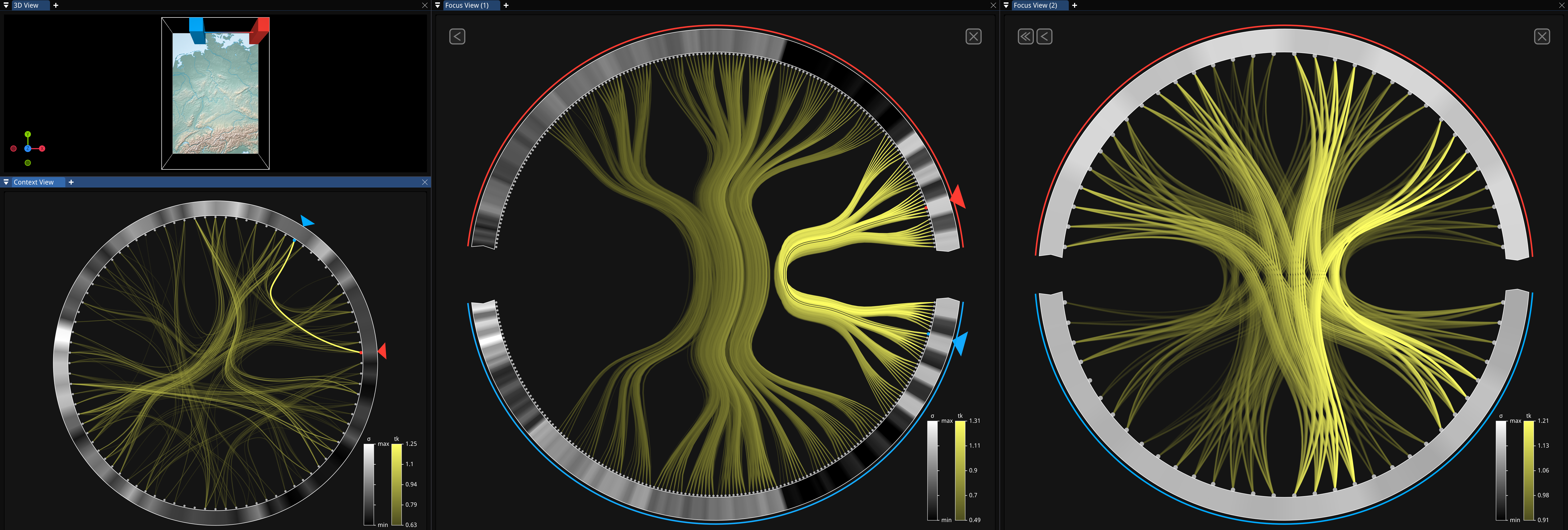}
\caption{%
Bottom left: Context view of a large weather forecast ensemble~\cite{Necker2020} shows mutual information (MI) maxima between regions in a 3D space partition, laid out on a chord along a space filling curve and visualized via edge bundling. The ensemble spread is encoded in the outer shell. Maxima are estimated via Bayesian optimal sampling (BOS) for all 3828 pairs of regions, each comprising $32 \times 32 \times 20$ grid points, in less than 16 seconds. Selected regions (red and blue triangle) are shown in a 3D view (top left).
Middle: Focus diagram shows refined MI estimates - computed on the fly using our GPU implementation in less than 4 seconds - between selected regions. Right: Focus view on regions selected in the first focus diagram. 
}
\label{fig:teaser}
\end{figure*}

\begin{abstract}
Visualizing spatial correlations in 3D ensembles is challenging due to the vast amounts of information that need to be conveyed. Memory and time constraints make it unfeasible to pre-compute and store the correlations between all pairs of domain points. We propose the embedding of adaptive correlation sampling into chord diagrams with hierarchical edge bundling to alleviate these constraints. Entities representing spatial regions are arranged along the circular chord layout via a space-filling curve, and Bayesian optimal sampling is used to efficiently estimate the maximum occurring correlation between any two points from different regions. Hierarchical edge bundling reduces visual clutter and emphasizes the major correlation structures. By selecting an edge, the user triggers a focus diagram in which only the two regions connected via this edge are refined and arranged in a specific way in a second chord layout. For visualizing correlations between two different variables, which are not symmetric anymore, we switch to showing a full correlation matrix. This avoids drawing the same edges twice with different correlation values. We introduce GPU implementations of both linear and non-linear correlation measures to further reduce the time that is required to generate the context and focus views, and to even enable the analysis of correlations in a 1000-member ensemble.
\end{abstract}


\section{Introduction}
\IEEEPARstart{
I}{n} this work, we use chord diagrams with hierarchical edge bundling \cite{HEB} for correlation visualization in 3D ensemble fields. We consider  
the correlations between the values of physical variables in the ensemble members at selected locations in 3D space. This will eventually enable us to analyze the used ensemble prediction system, which generates the ensemble members using variations of the initial simulation conditions and model formulations, by conveying regions where these variations result in similar or dissimilar relative behavior of the member values. Notably, by interpreting a single ensemble member at $T$ distinct time steps as an ensemble comprising $T$ members, our approach can be immediately applied to convey temporal correlations in a single field.

When using chord diagrams, however, the space along the circular layout is limited and large numbers of entities cannot be well represented. While a chord diagram can at most show few hundreds of entities and their relationships, considerably more need to be shown to convey all point-to-point correlations in a 3D field.  

To address this limitation, we propose using 
two chord diagrams which are shown side by side. The context view (see left of \cref{fig:teaser}) uses a chord diagram that shows the correlations between all regions in the domain that emerge by partitioning space into disjoint sub-domains. The focus view (see middle and right of \cref{fig:teaser}) shows refined correlations between exactly two selected regions. The user controls the refinement by selecting edges in the chord diagrams and can refine down to the level where point-to-point correlations are shown. 

The correlation between two regions is indicated by the maximum of all point-to-point correlations between pairs of grid points in either region. In general, however, 
computing all point-to-point correlations between a pair of regions---and for all pairs shown in the context view---is unfeasible due to excessive computing times. I.e., for the ensembles we consider in this work, the computation of all point-to-point correlations using the Pearson Product-Moment Correlation Coefficient (PPMCC) as a linear dependence measure takes more than two days on a recent GPU, and the computation of more complex measures like the non-linear Mutual Information (MI) takes considerably longer. Storing all point-to-point correlations, on the other hand, requires more than 2.8 TiB of memory (assuming two bytes per correlation value) for a single variable and, thus, is unfeasible too. 

Our strategy to reduce the time for generating the context and focus diagrams is twofold. Firstly, we provide efficient GPU implementations---especially for computing MI---to speed up the computation of single correlation samples. Secondly, and inspired by the approach of Chen et al.~\cite{Chen2011}, we employ importance-based correlation sampling to significantly reduce the number of point-to-point correlations that need to be computed.
However, we use Bayesian Optimal Sampling (BOS) to automatically select the set of point-to-point correlations that provide a good estimate of the maximum correlation between any two points from different regions. Especially the time required for generating the context view, where correlations between many pairs of large regions need to be estimated, can be reduced significantly via BOS. 
If the data does not fit into GPU memory, the system switches to an aggregate data representation using statistical means, at a resolution level that can be stored on the GPU. BOS is then performed on this level to avoid performance losses, at the expense of higher uncertainty in the maximum correlation estimates. 

Our main contributions are:

\begin{itemize}
    \item Locally adaptive BOS is used to compute the maximum point-to-point correlation between two regions from only few correlation samples.
    \item A novel GPU implementation of point-to-point MI computations exploiting parallel search and sorting operations.
    \item Entities, i.e., domain regions or points, are arranged along the circular chord layout using a space-filling curve.
    \item A focus diagram shows a selected pair of regions opposite to each other on the circular layout.
\end{itemize}

Chord diagrams with hierarchical edge bundling have been selected for correlation visualization because they can effectively reduce the amount of displayed edges, and correlations below a user-selected threshold can simply be omitted. In correlation matrices, as an alternative, entries corresponding to low values can be left blank, but they are shown nevertheless and take up space. Thus, chord diagrams can more effectively reduce visual clutter.
Furthermore, when all correlations for a selected entity should be analyzed using a correlation matrix, the user follows the corresponding row or column. In this case, only showing the upper or lower triangular matrix requires the user to change ``direction''. Thus, the full matrix including redundant information needs to be shown, which is avoided when using a chord diagram.
However, when correlations between two points and different variables are analyzed, it matters at which point which variable is considered. The same edge then needs to be drawn twice with different correlation strength in a chord diagram, leading to visual clutter. To avoid this, we show these correlations in a correlation matrix, which is then completely filled. 
Adaptive refinement is supported as described, by selecting matrix elements and showing the refined regions in a focus diagram using again a correlation matrix. 


The remainder of this paper is structured as follows. After reviewing previous work, we describe and evaluate BOS for estimating the maximum correlation between two regions using a reduced set of correlation samples, and we introduce a novel GPU method for computing the MI between many pairs of random variables in parallel. Next, we introduce focus+context chord diagrams for correlation analysis in 3D ensemble fields. 
We then perform a quality and performance evaluation of the proposed approach, and demonstrate the use of the proposed approach with a synthetic and two large 3D weather forecast ensembles.
The paper is concluded with ideas for future work.





\section{Related Work}\label{sec:related-work}



In this work, we address the problem of how to efficiently visualize spatial correlations in large 3D ensembles of scalar fields. This is in contrast to many previous approaches for ensemble visualization, which have often been devoted to the visual analysis of the ensemble spread, using feature- and location-based approaches~\cite{met3d,FutureEnsChallenges,EnsSurvey}. 
This includes approaches that provide visual abstractions of the major trends in ensembles of line or surface features~\cite{EnsemblesCentralTendency,EnsemblesIsoContours,  ApproachesUncertaintyVisualization,NoodlesEnsembleUncertainty,ContourBoxplots}, as well as location-based approaches that visualize local statistical data summaries~\cite{Pfaffelmoser-GradUnc-2013,EnsemblesCentralTendency,EnsemblesIsoContours,ApproachesUncertaintyVisualization} or find compact representations of ensemble data ~\cite{hazarika-2018}.

\subsection{Ensemble Correlation Analysis}

The analysis of spatial (auto)correlations, concerned with both time-varying and ensemble data, is an important task in meteorology and climatology. 
Nocke et al.~\cite{npg-22-545-2015} introduce concepts from visual network analytics to analyze climate networks. In particular, they use edge bundling in geo-referenced networks to indicate spatial correlations in 3D climate data. Wilks~\cite{MetCorr} discuss the effect of spatial correlations among the grid points on statistical significance tests in atmospheric sciences. They focus on the analysis of spatial correlations of linear trends in annual precipitation, and plot the correlations of pairs of cells with a certain maximum distance threshold as a scatter plot and fit a decay function to show the decrease of the correlation by the cell distance. For iso-contours in scalar ensemble fields, Ferstl et al.~\cite{EnsemblesIsoContours} assess the spatial correlation of their occurrence at different locations in the domain. To counteract the high computational cost of dependence measures like MI, Farokhmanesh et al.~\cite{NeuralFieldsStatDepEnsembles} propose the use of neural dependence fields (NDFs) to reconstruct approximated dependence fields at runtime from a compressed form for interactive visualizations. The underlying neural network is trained using sparse, randomly selected correlation samples.
Global teleconnections are visualized by Delalene et al.~\cite{esd-14-17-2023} to analyze in interannual to decadal climate variability. They use dependence measures to infer spatial functional networks between clustered sub-domains, in combination with correlation matrices to visualize such dependencies. Kumpf et al.~\cite{kumpf} use correlation clustering to assess the sensitivity of numerical weather forecast quantities to changes in model variables. Their approach aims to reveal the confidence in the sensitivities obtained via ensemble sensitivity analysis \cite{Ancell-2007,Torn-2008}, which is used in meteorology to determine the origin of forecast errors and improve placing of targeted observations. Evers et al.~\cite{10.2312:envirvis.20231108} introduce the use of multi-dimensional embeddings for determining clusters in time series correlation in 2D multi-field climate ensembles. 

Besides linear correlation measures such as the Pearson correlation coefficient, non-linear measures like MI have been considered in atmospheric and climate sciences. 
Laarne et al.~\cite{LaarneMI} use MI for exploring non-linear dependencies in atmospheric data. They use it both for analyzing temporal autocorrelations of a single field using line plots and correlations between multiple fields using static correlation matrices.
Babel et al.~\cite{BabelMI} use MI for selecting the most suitable explanatory variables that are fed to a neural network for rainfall forecasting. Ning et al.~\cite{NingMI} use MI for uncertainty assessment in precipitation forecasts.

\subsection{Correlation Visualization}


In particular for the analysis of multivariate data, where multiple variables are given at each data point, a number of approaches for correlation visualization have been proposed. The survey by He et al.~\cite{HeSurvey2019} gives an overview of the different research areas in the field of multivariate spatial data analysis. One line of research is dedicated to the finding of appropriate similarity measures to support multivariate correlation analysis, such as gradient-based measures as introduced by Sauber et al.~\cite{Sauber2006}, Gosink et al.~\cite{Gosink2007} and Nagaraj et al.~\cite{Nagaraj2011}. Zhang et al.~\cite{Zhang2015} introduce correlation maps, which show pairwise correlation between variables in a graph, and combine them with a parallel coordinate view indicating the main correlation structures on the data level. Liu and Shen~\cite{Liu2016} introduce probabilistic association graphs for the analysis of informativeness and uniqueness of multivariate data. Like in our work, their visual workflow provides a circular chord diagram for visualizing the relationships between pairs of entities. In some previous works ~\cite{Biswas2013,Wang2018,Berenjkoub2019}, MI has been considered as a measure of statistical dependence between different variables.


Another popular approach for correlation analysis is correlation clustering. Pfaffelmoser and Westermann~\cite{Pfaffelmoser2012} cluster regions of 2D scalar fields based on a measure of the correlation of a random variable to its surrounding, called the correlation neighborhood. Liebmann et al.~\cite{Liebmann2018} use hierarchical correlation clustering in combination with a dendrogram visualization. Berenjkoub et al.~\cite{Berenjkoub2019} combine the segmentation of the spatial domain based on correlation strength with according coloring of pathlines, yet they restrict to correlations among pairs of attributes at narrow spatial regions and fixed locations over time or along pathlines. Evers et al.~\cite{Evers2021} combines a hierarchical correlation segmentation algorithm with correlation heat maps. Biswas et al.~\cite{Biswas2013} cluster variables with similar distributions into groups and select a representative of each cluster. Variables are displayed in a diagram using a force-directed graph layout, where MI between two variables is used for computing the attractive force between two nodes. This graph view is linked with parallel coordinate plots and an isosurface view for further exploration. Due to its computational complexity, correlation clustering has been restricted to 2D fields so far.

\begin{figure*}[ht]
\centering
\includegraphics[height=4cm]{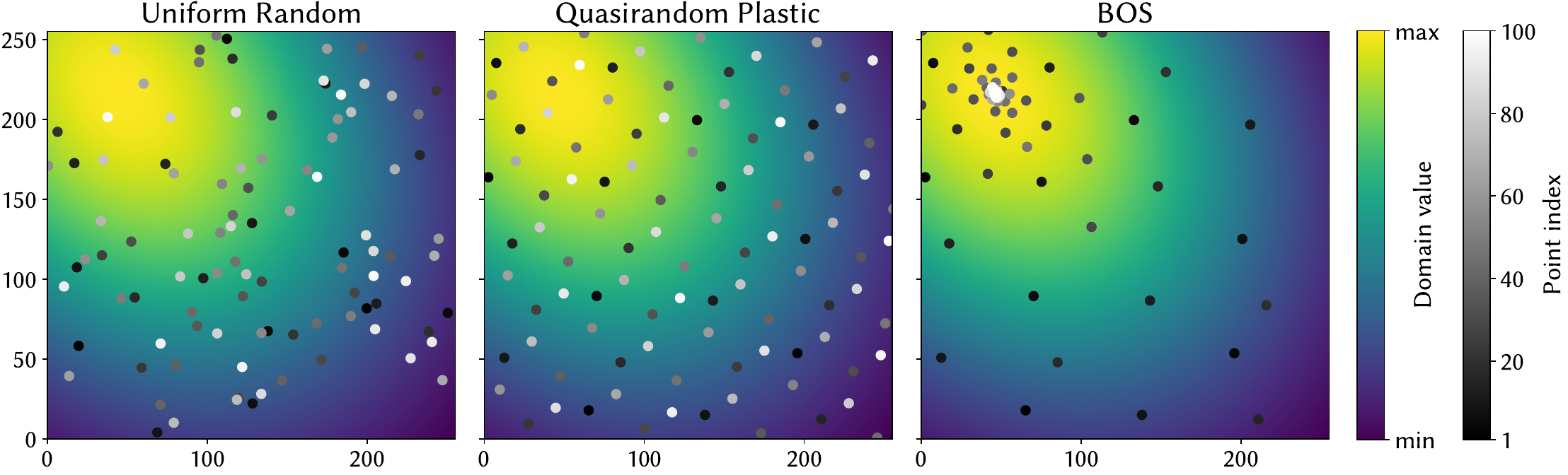}
\caption{Locations of the first 100 samples picked by different sampling schemes for finding the maximum in a 2D field, here the probability density function of a 2D multivariate normal distribution.}
\label{fig:sampling-strategies}
\end{figure*}

To address the computational complexity of computing correlations, i.e., quadratic complexity in the number of entities to put into relation to each other, Su et al.~\cite{Su2014} describe a system for parallel and distributed correlation analysis. The method generates a joint bin representation as a data aggregate on which measures like the MI can then be estimated efficiently.  Chen et al.~\cite{Chen2011} propose using domain knowledge to first draw a set of samples from a simulation grid and compute the correlation volumes for each of these samples. A correlation volume stores the correlations from all points in the grid to a reference point. A matrix containing the distances of each pair of correlation volumes is then calculated and used to decide where more samples need to be computed. 


\subsection{Linearization of Spatial Data}
Diagram techniques provide effective encodings of information in the form of schematic representations. When encoding the elements of a spatial domain like a grid in a diagram, the elements often need to be linearized to lay them out along curves or surfaces. When working with 2D or 3D spatial data, space-filling curves are often used to generate such a layout, i.e., a mapping of multi-dimensional positions to a one-dimensional index, under the objective that spatial proximity is preserved. Popular choices are Z-order curves or Peano-Hilbert curves \cite{Hilbert}. Zhou et al.~\cite{SpaceFillingCurves} introduce data-driven space-filling curves, which take into account the underlying data to increase the spatial proximity of elements with similar values. After linearization of the spatial elements, relationships between elements can be represented by connecting the linearized elements with lines. When displaying a lot of lines at once, the visualization can become cluttered due to many crossing lines and structures in the data become hard to perceive. Hierarchical edge bundling~\cite{HEB} partly alleviates this limitation by visually bundling lines based on a hierarchical structuring of the elements. In our work, we use a Z-order curve for linearizing the vertices of a 3D simulation grid. The advantage of Z-order curves over the data-driven curves by Zhou et al.~\cite{SpaceFillingCurves} is that one can create an octree hierarchy over the elements, which can then be used for hierarchical edge bundling and on-demand subdivsion of the displayed data level.
A popular example demonstrating the use of diagram techniques with linearization of spatial elements is the work by Demir et al.~\cite{Demir2014}. After linearization, they use a histogram-based diagram view combining line and bar charts. Wang et al.~\cite{Wang2011} introduce a circular graph that can show the information transfer between pairs of variables. Recently, Weissenbock et al.~\cite{Weissenbock2019DynamicVL} extend on the work by Demir et al.\ and propose to adaptively scale the mapping that is obtained via a space-filling curve, to enable focusing on important spatial regions. 

Chord diagrams have been used in a number of previous works for the analysis of hierarchical data coupled with focus+context-like approaches. Rees et al.~\cite{ChordDiagInteraction} propose a technique for the interaction with large chord diagrams, proposing brushing and deformed chord diagrams acting as focus views. Bae and Lee~\cite{TagRelationsFC} combine chord diagrams with magnifying lens-like focus views for the analysis of the relation of tags. Gou and Zhang~\cite{TreeNetViz} describe a technique for the multiscale exploration of network data organized in a tree, where aggregate elements can successively be refined in the same diagram view.


\section{Correlation Sampling}\label{sec:bayesian}

We assume that a simulation ensemble with $E$ members 
is given on a 3D regular grid of size $X \times Y \times Z$.
At every grid point, one or more variables are given per ensemble member. We further assume that $M$ different entities can be encoded along the circular layout of a chord diagram, where $M$ needs to be selected depending on the resolution of the diagram.
Then, the Cartesian grid is partitioned into $M$ sub-grids of size $C = X_d \times Y_d \times Z_d$. These sub-grids are subsequently called bricks. Partitions are built so that the maximum difference between any pair of elements from $\{X_d, Y_d, Z_d\}$ is minimized, i.e., bricks are mostly isotropic in the three dimensions. In the same way, each brick is recursively refined into $M/2$ bricks of ever smaller size, so that all siblings of exactly two bricks, i.e., the pair for which correlations are visualized, can be arranged along the circular chord layout. The refinement process is recursively repeated until bricks are comprised of one single data value. 

To determine an indicator of the correlation between two bricks, point-to-point correlations between pairs of grid points in either brick are computed, and the maximum of these correlations is used. 
In the following, we describe the importance-based sampling strategy we employ to obtain a good estimator of these maxima from only a low number of effectively computed point-to-point correlations. 


\subsection{Bayesian Optimal Sampling}\label{sec:bos}
One approach for finding the maximum correlation between a pair of bricks is random sampling of correlations at grid points in each pair of bricks. 
One sample position corresponds to a position in a 6 dimensional space, where 3 dimensions represent the position in one brick, and the remaining 3 dimensions represent the position in the respective other brick.
We have experimented with uniform random sampling and quasi-random low-discrepancy sequences, like Halton~\cite{Halton1964} and plastic~\cite{PlasticSequence} sequences.
Previous works on the use of low-discrepancy sampling in quasi-Monte Carlo numerical integration~\cite{QMC} have shown superior convergence properties of quasi-random low-discrepancy sequences over random sampling. Unlike the case of Monte Carlo integration, we were not able to perceive any significant differences in the convergence properties of uniform pseudo-random and quasi-random low-discrepancy sampling strategies for stochastic optimization. We have summarized our experiments in \cref{sec:bayopt-eval}.
However, a disadvantage of both random and quasi-random maximum sampling is that local or global maxima can be missed. This raises the question whether available samples can be used to infer in which regions new samples would be most likely to result in improved maxima.

Bayesian optimization~\cite{bayesopt_1989,bayesoptbook_2023} aims to solve the optimization problem $\max_{\theta \in D} f(\theta)$ for a blackbox function $f$ and domain $D$, which may be hard to compute.
Underlying the optimization is the concept to minimize the number of function evaluations, i.e., the number of correlation samples that need to be computed in our scenario. The basic ingredients of the optimization are the probabilistic surrogate model and the acquisition function. While the former is used to express Bayesian belief about the outcome of the objective function that is derived from known evaluations, the acquisition function chooses the next sample to evaluate. \cref{fig:sampling-strategies} compares the specific sampling patterns used by (quasi\=/)random sampling and BOS for finding the maximum value in a 2D domain.

Specifically, Gaussian process surrogate models can be used to model a distribution of the function values for each location in the parameter space. 
To restrict the distribution of available functions to those that agree with the available correlation samples, the model is conditioned on the new sample and the given set of samples. We randomly sample an initial set of correlations and update the model accordingly. The acquisition function is then maximized with respect to the selected model, to predict the locations of the next correlation samples. For the acquisition function, we use the Upper Confidence Bound (UCB)~\cite{UCB}, and for optimizing it, a randomized version of DIRECT-L~\cite{directl} is used, which is provided via a binding to the NLopt library~\cite{nlopt}. We have evaluated all gradient-free optimization algorithms provided by NLopt and found this optimizer to give the best error metric results in our concrete settings, see \cref{sec:bayopt-eval}. The Matern kernel~\cite{matern} is used for the covariance function of the underlying Gaussian processes.

Most Bayesian optimization algorithms operate on continuous search spaces. Daulton et al.~\cite{BayOptDiscrete} describe how to use probabilistic reparameterization, where the discrete search space of voxel center pairs is considered. As proposed by Daulton et al., we map continuous search space parameters $\theta \in [0, X_d - 1] \times \cdots \times [0, Z_d - 1]$ to discrete 6D grid positions $P_i = \lfloor \theta_i \rfloor + B_i, B_i \sim \textrm{Ber}(\theta_i - \lfloor \theta_i \rfloor)$. $\textrm{Ber}(p)$ is the Bernoulli distribution, which has an outcome of 1 with a probability of p and an outcome of 0 otherwise. Consequently, the continuous positions between voxel centers will be mapped to closer centers with higher probability. To obtain support for Gaussian processes and optimization, we integrate the library Limbo~\cite{Limbo} into our application, and batch the acquisition of new correlation samples for multiple brick pairs using multiple CPU threads. This enables us to efficiently use GPU correlation computations in conjunction with Bayesian optimization on the CPU. Parallel executions on the GPU are exploited to compute correlations for multiple feature vectors simultaneously. 

\subsection{GPU-Accelerated Correlation Computation}

Since the proposed workflow builds upon in-turn computations of large sets of point-to-point measures of statistical dependence, these computations need to be highly efficient to not hinder interactivity. In order to not restrict the workflow to computationally lightweight measures like PPMCC, a novel optimized GPU implementation for computing MI has been developed. 

Cover and Thomas~\cite{MI} define MI as the ``relative entropy between the joint distribution and the product distribution'' of two random variables, i.e., 
\begin{equation}
    \label{eq:mi}
    MI(X;Y) = H(X) - H(X|Y) = H(Y) - H(Y|X),
\end{equation}
were $X$ and $Y$ are two random variables, and $H$ is the entropy. Like PPMCC, also MI is symmetric in the two random variables, yet it is in the range $[0, \infty)$ and does not distinguish between negative and positive correlations.

While PPMCC requires merely to compute means and variances and can, thus, be realized efficiently on the GPU, MI requires to estimate the joint discrete probability density function from the joint histogram of the two variables in order to compute the conditional entropy.  
A popular approach for computing the joint histograms is via binning, i.e., the two continuous random variables are discretized and a joint and two marginalized histograms are computed for the discretized realizations. MI is then estimated from the discretized realizations. Binning has been both parallelized and ported to the GPU in previous works, e.g., for the purpose of correlation analysis~\cite{Su2014} and medical image registration~\cite{MIMed2008,MIRegistr2010A,MIRegistr2010B,MIImgAlign2022}.

As has been demonstrated and analyzed by Kraskov et al.~\cite{KraskovMI}, however, binning introduces systematic errors and result in low fidelity estimators. To avoid this, they propose using entropy estimates that are derived from the Chebyshev distances between the k-th nearest neighbor in the joint distribution space. Given $X$ and $Y$, the joint distribution $Z = (X,Y)$ is represented by joint 2D data samples $z_i = (x_i, y_i)$.
For each joint sample $z_i$, its distance $\epsilon_i$ to the $k$-th nearest neighbor is calculated using the Chebyshev distance, i.e., for a k-th nearest neighbor $z_j$, $d(z_i, z_j) = \max\{\lvert x_i - x_j \rvert, \lvert y_i - y_j \rvert\}$. We choose $k = \lceil \frac{3n}{100} \rceil$ in accordance to the results by Kraskov et al.\ for correlated Gaussian distributions.
Then, for a joint sample $z_i$ the numbers $n_{x,i}$ and $n_{y,i}$ of joint samples fulfilling respectively $\lvert x_i - x_j \rvert < \epsilon_i$ and $\lvert y_i - y_j \rvert < \epsilon_i$ are computed, and the MI is estimated as $\widetilde{MI}(X,Y) = \psi(n) + \psi(k) - \frac{1}{n} \sum_{i=1}^{n} \psi(n_{x,i}) + \psi(n_{y,i})$. Here, $n$ is the number of realizations of the random variables, i.e., the ensemble member count $E$ in our application, and the so-called digamma function $\psi(z)$ is the logarithmic derivative of the gamma function $\Gamma(z)$~\cite{digamma}, i.e.,
\begin{equation}
    \label{eq:digamma}
    \psi(z) = \frac{d}{d z} \ln \Gamma(z) = \frac{\Gamma'(z)}{\Gamma(z)}.
\end{equation}

Notably, due to the complexity of k-nearest neighbor search and frequent evaluations of the digamma function, only roughly 5000 point-to-point MI samples can be computed per second on the CPU for the largest of our data sets, compared to approximately 800000 per second using PPMCC (cf.~\cref{tab:perf}). Since we are not aware of any GPU implementation of the Kraskov estimator (abbreviated as KMI in the following sections),
we introduce such an implementation in the following. This implementation achieves a speed-up of a factor of 10, and thus gives the speed that is required for an interactive correlation analysis as intended by our approach 



At the core of an efficient computation of KMI is a data structure that allows to efficiently perform k-nearest neighbor queries for many joint samples in parallel on the GPU. Notably, the search data structure needs to be rebuilt for each pair of points for which MI is estimated, i.e., for each set of $E$ joint data samples $z_i = (x_i, y_i), i \in \{1, \dots, E\}$. Then, the data structure is used to determine the distance to the $k$-th nearest neighbor for every joint sample.
In principle, k-d trees
are well suited for this task, yet recursive tree traversal and heterogeneous code paths for different data samples in the tree construction and distance computation make an efficient GPU implementation challenging. 
Since the GPU is based on a single instruction, multiple threads (SIMT) model, where many threads are run in lockstep, it can only sequentially evaluate different code branches for a group of threads. Furthermore, to reduce latencies, 
as many memory access operations as possible should be moved outside of branches. Lastly, GPU shading languages usually do not support recursion, which is used in most available k-d tree implementations. 

In a number of previous works, these restrictions have been addressed with specific adaptations of k-d tree construction and traversal on the GPU~\cite{ParallelKdTreeBuild}. The use cases of these implementations, however, differ significantly from ours. While usually a single k-d tree representing an entire scene is built once and many queries are executed in parallel, in our scenario as many k-d trees as MI values need to be built, i.e., one k-d tree per pair of joint data samples. Then, each tree is used to accelerate the $E$ k-th nearest neighbor searches for each joint sample $z_i$.
Thus, in our implementation we assign the computation of one point-to-point MI value to one compute thread, so that a single thread computes its specific k-d tree and performs all k-th neighbor searches sequentially. As neighboring threads of a thread group will build and traverse k-d trees based on different data in lockstep, it is important to avoid divergence caused by the different data being processed per thread.

One of the most recent GPU-friendly k-d tree traversal algorithms by Wald~\cite{wald2022stackfree} employs iterative stack-free tree traversal to avoid recursion and minimize diverging code branches. Compared to a custom iterative k-d tree traversal using a manually managed stack as a replacement for recursion, the approach turned out to be about $15\%$ faster for the large ensembles we consider in this work. However, when using the stack-free implementation we noticed that the fairly heterogeneous sets of data values assigned to the threads in one thread group can lead to significant execution divergence between neighboring threads during the sorting operation required for k-d tree construction. To reduce this effect, we have incorporated the GPU heap-sort implementation by Kern et al.~\cite{Kern:2020:TVCG}. Instead of using 
variable loop bounds and branching to avoid unnecessary element swap operations like sorting algorithms tailored to CPUs like quicksort~\cite{quicksort},
the proposed heap sort implementation accepts potentially unnecessary element swaps in favor of avoiding branching. By this, the computation of the MI is further accelerated by about $10\%$ to $20\%$ (depending on the ensemble size) compared to using an adapted, iterative version of introsort~\cite{introsort}. Introsort is a hybrid sorting algorithm switching from quicksort to another sorting algorithm when the data recursively sorted becomes too small for quicksort to work efficiently. Introsort is used, for example, by the GNU standard C++ library~\cite{gccsort}.

Finally, we use the Lanczos approximation \cite{Lanczos,LanczosGamma} in our implementation to efficiently evaluate the digamma function on the GPU in constant time. The Lanczos approximation is a method for the numeric approximation of the gamma function and can also be extended to the digamma function using the relationship between the two functions in \cref{eq:digamma}.
In \cref{sec:perf}, we compare the performance of our GPU implementation of KMI to an optimized implementation on the CPU, and observe speed-ups of approximately one order of magnitude.


\subsection{Sampling from Spatial Aggregates}\label{sec:aggregate-sampling}
If the ensemble cannot be stored in GPU memory, sampling point-to-point correlations in the initial data becomes far more time consuming, since it requires streaming the entire ensemble when new samples are needed. Notably, however, for realistically sized data sets this affects only the generation of the context view, where many point-to-point correlations need to be computed for each pair of initial bricks. Once the user has selected a brick pair for analysis in the focus view, the initial data values of both bricks are quickly streamed to the GPU and BOS is performed as described.

To avoid the aforementioned restriction,
a hierarchical ensemble representation using the same partitioning strategy as for brick refinement is computed first. 
We have evaluated both a mean-tree and a max-tree as alternatives. 
The mean/max-tree stores for each brick and each ensemble member---starting at the first refinement level---the means/maxima of all variables at the grid points represented by this brick. For the largest ensemble we consider in this work, this process takes about 12 seconds on our target architecture per variable. For computing the correlation with respect to a selected variable between a pair of bricks, BOS is then performed using the mean/maximum values at the highest resolution tree-level that just fits into GPU memory. 

It is clear that using aggregate values for correlation estimation introduces errors in the estimated correlation maxima, because significant and in particular isolated high correlations can be overlooked. On the other hand, our experiments show that performing maximum correlation estimation on pre-computed means gives results that are close to those obtained via BOS and sampling point-to-point correlations from the initial data. Furthermore, when showing the ensemble spread in the chord diagrams as described below, this information indicates those bricks where the estimation is uncertain and further refinements should be performed. As our experiments have indicated that both the mean- and max-tree give more or less equal accuracy, we have decided to support the mean-tree in our implementation.

\begin{figure}[t]
\centering
\includegraphics[height=5cm]{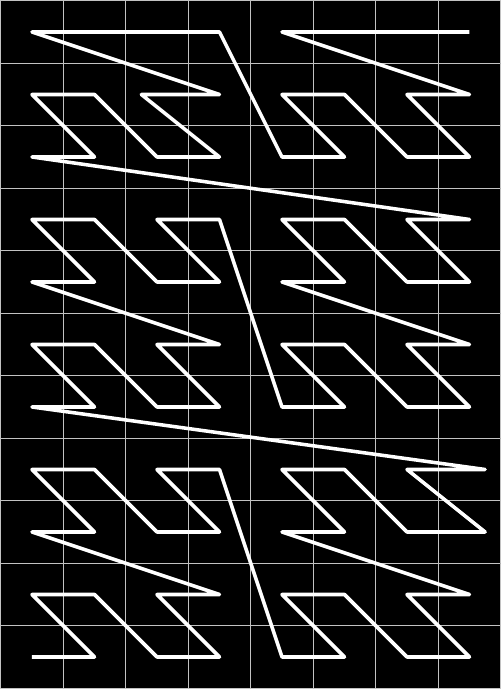}
\includegraphics[height=5cm]{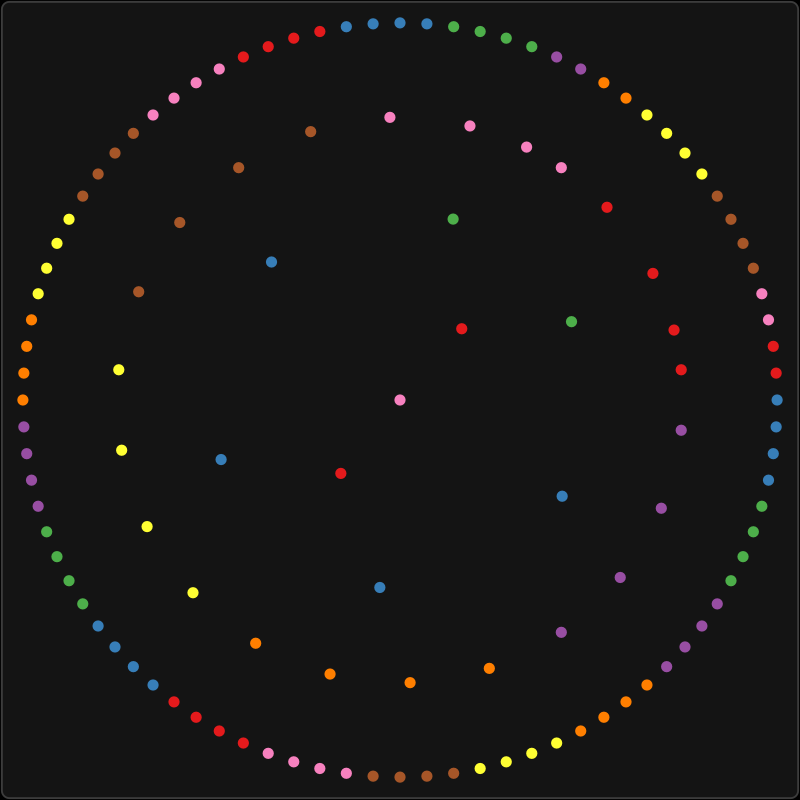}
\caption{Left: Subdivision of a $250 \times 352 \times 20$ simulation grid into $8 \times 11 \times 1$ bricks of size $32 \times 32 \times 20$. The bricks are traversed using a Z-order curve. Right: Nodes in the chord diagram. The leaves ordered by the Z-order curve are linearized in the outermost circle. Inner octree nodes used for hierarchical edge bundling (they are hidden during rendering). Nodes with the same parent are assigned the same color.}
\label{fig:zorder}
\end{figure}

\section{Chord Diagrams for 3D Fields}\label{sec:chord-diagrams}

As discussed above, the ensemble domain is partitioned into $M$ regions, and each region is again partitioned into $M/2$ sub-regions. $M$ is the number of entities that can be encoded along the circular layout of a chord diagram, and it needs to be selected depending on the resolution of the diagram.
For instance, when plotting a chord diagram on a 512x512 viewport, we typically choose $M=128$ to approximately have 4-7 pixels between each entity on the circle so that edges between entities can be well distinguished.

Bricks are laid out along the circle through a Z-order curve (see \cref{fig:zorder}~left), so that the spatial proximity of bricks is preserved as good as possible in the circular layout. Unlike other space-filling curves, like the data-driven approach by Zhou et al.~\cite{SpaceFillingCurves}, Z-order curves have the property that when the data is hierarchically partitioned using an octree structure, all siblings of a node are visited by the space-filling curve before proceeding to the next parent node. Since we use such an octree layout for brick refinement, and also build an octree-based partition above the initial $M$ bricks in turn, we can effectively make use of hierarchical edge bundling~\cite{HEB}, i.e., bundling lines based on a hierarchical structuring of the entities on the circular layout. By using a linearization of bricks along a Z-order curve, we directly obtain a hierarchy for the bricks that are visualized in the chord diagram. The root node of the octree is placed in the center of the chord diagram and the inner nodes are positioned from the center to the outer perimeter of the diagram. The angle of an inner node is the average of the angles of their child nodes. This hierarchical arrangement is demonstrated in \cref{fig:zorder}~right. Finally, as proposed by Holten~\cite{HEB}, when connecting two entities (i.e., leaf nodes of the octree) the inner nodes along the shortest path between the two leaves are used as control points for constructing a smooth B-spline curve, i.e., an edge. B-splines are discretized using the de Boor algorithm~\cite{deBoor1971} and rendered using the vector graphics library NanoVG~\cite{NanoVG}.

\begin{figure*}[t]
\centering
\includegraphics[height=5.05cm]{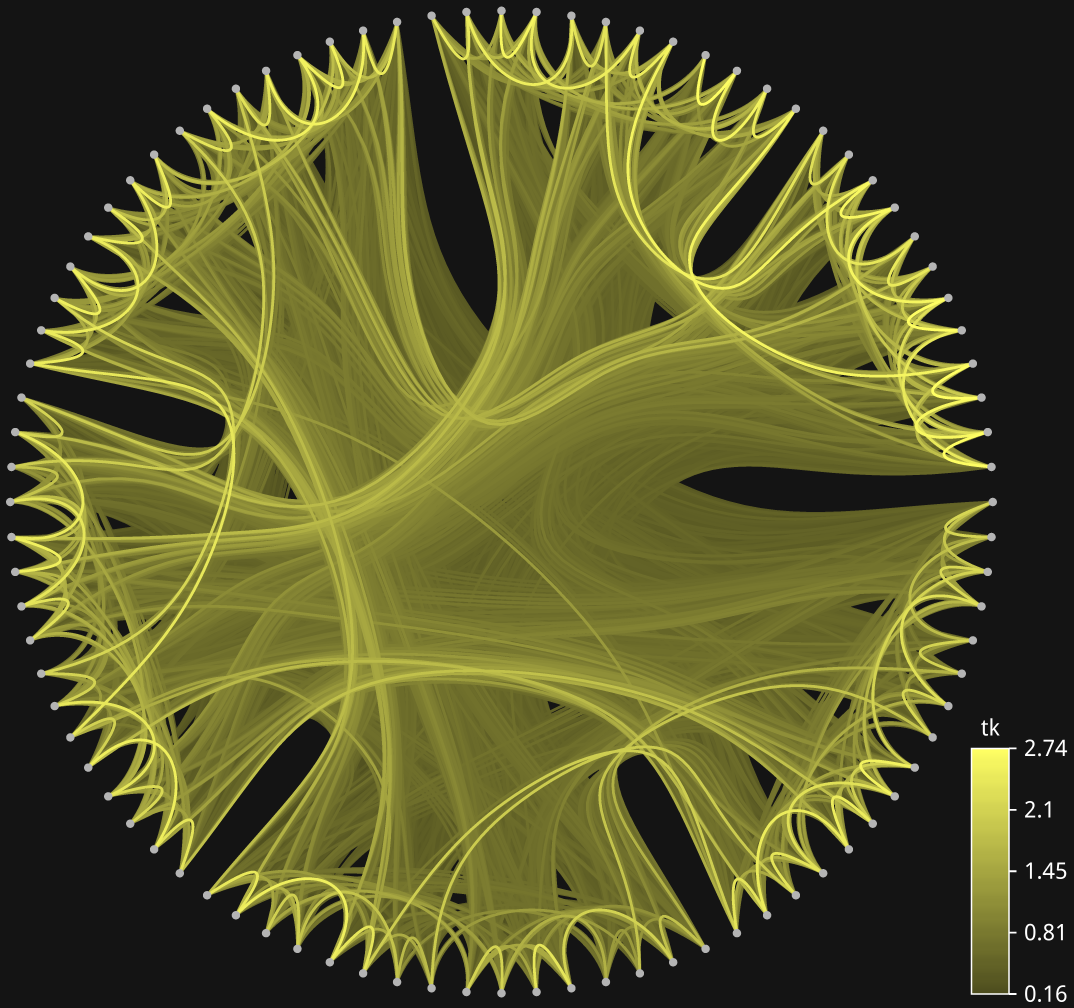}
\includegraphics[height=5.05cm]{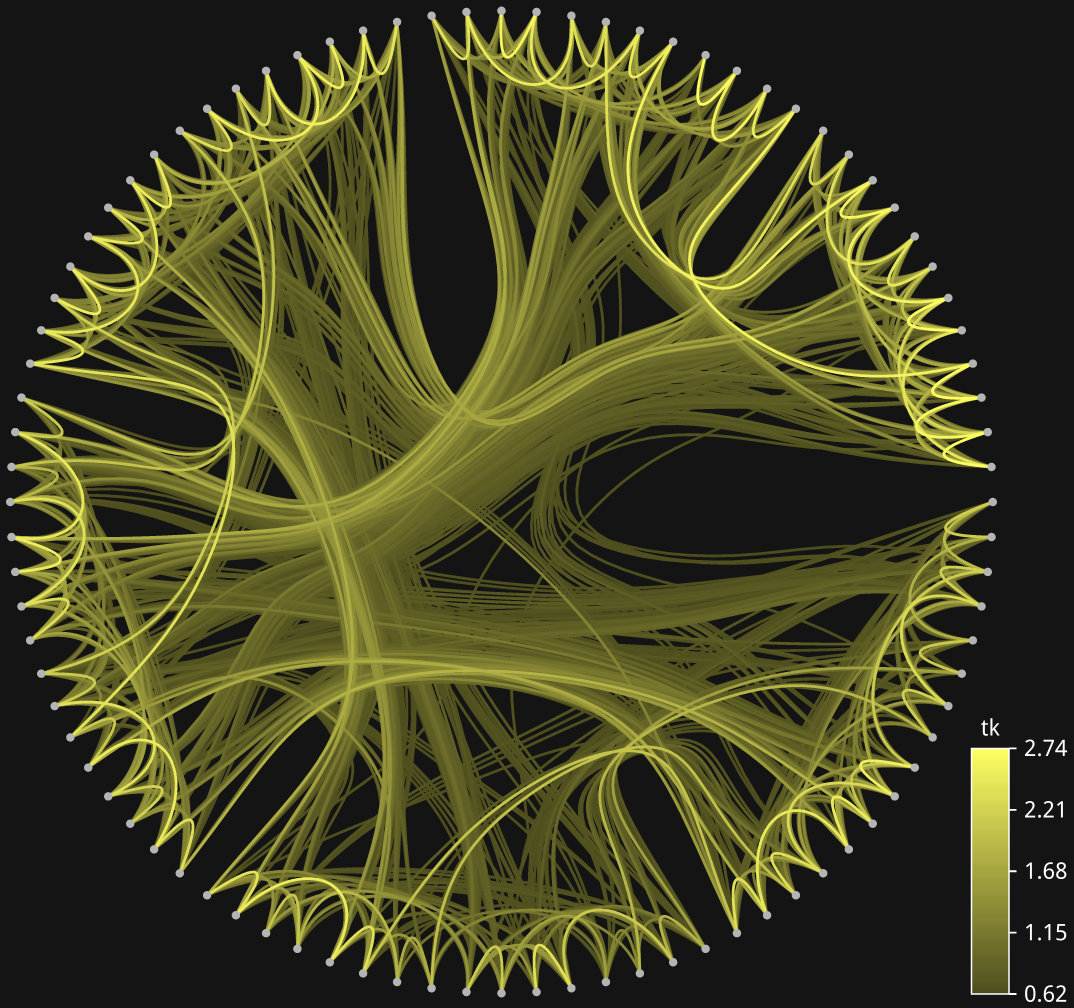}
\includegraphics[height=5.05cm]{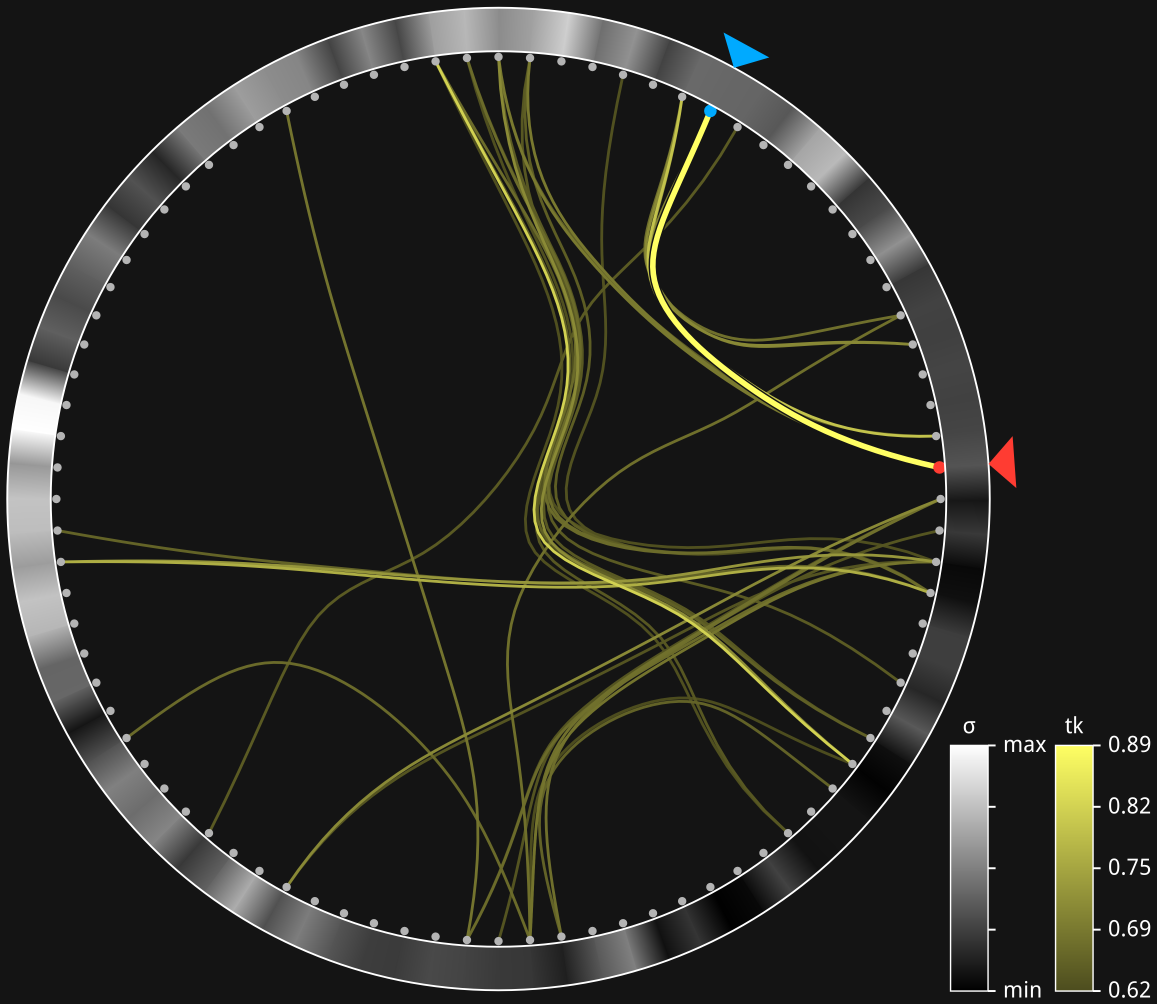}
\caption{Context chord diagram using MI as dependence measure for a large simulation ensemble~\cite{Necker2020}. Left: Spatial correlations in the temperature field \texttt{tk}. Middle: Sub-selection of correlations with MI $\ge 0.62$.
Right: Sub-selection of correlations with brick distance $\ge 570$km (measured from the brick centers) and MI $\ge 0.62$.
Ensemble spread $\sigma$ of \texttt{tk} is shown on the outer ring.}
\label{fig:context-chord}
\end{figure*}

\subsection{Context Chord Diagram}

In the context diagram, the $M$ bricks, each representing a sub-grid of size $X_d \times Y_d \times Z_d$, are aligned along the circular chord layout. They are visualized via small circles and connected by edges indicating their pair-wise correlations. Each edge represents the maximum correlation between any of the grid points in two connected bricks (see \cref{fig:context-chord}).
The user can select showing correlations in a selected range, to emphasize important relationships. Furthermore, only correlations between bricks in a certain distance range can be shown, which enables to switch between long-, mid- and short-range correlations. 
Edges are ordered with respect to increasing correlation magnitude, to also emphasize strong negative correlations, and rendered in this order on top of each other. Correlation strength is mapped to color, starting at the background color (low strength) and fading towards a selected key color (high strength).
The user can specify a range of correlation values to use for filtering the displayed edges. By default, the selected range is also used as the range for color mapping.

As shown in \cref{fig:context-chord}~right, the outer ring around the circular layout is used to display additional ensemble information via a greyscale colormap, 
i.e., the per brick ensemble spread (average of the standard deviation $\sigma$). Since high ensemble spread indicates a high uncertainty in the ensemble simulation, the spread hints at interesting regions for further analysis and, in particular, indicates the uncertainty when estimating brick-to-brick correlations using mean values over successively refined sub-regions.

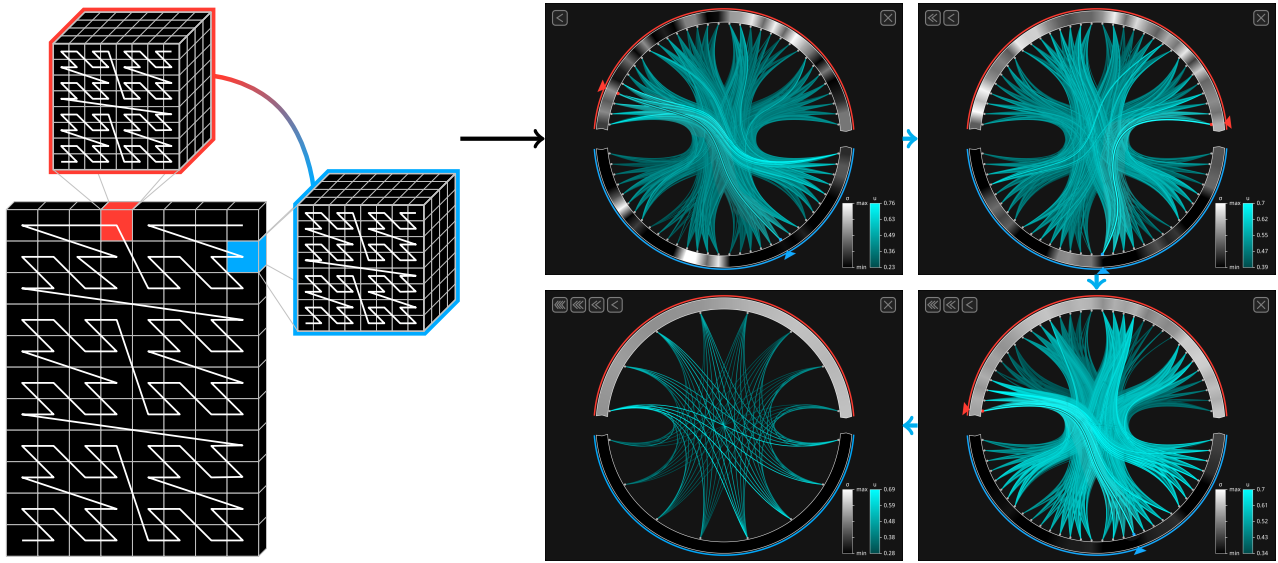
\begin{figure*}[ht]
\centering
\begin{tikzpicture}

\begin{scope}[node distance=0.2cm]
  \node [inner sep=0pt] (a) {\includegraphics[height=3.6cm]{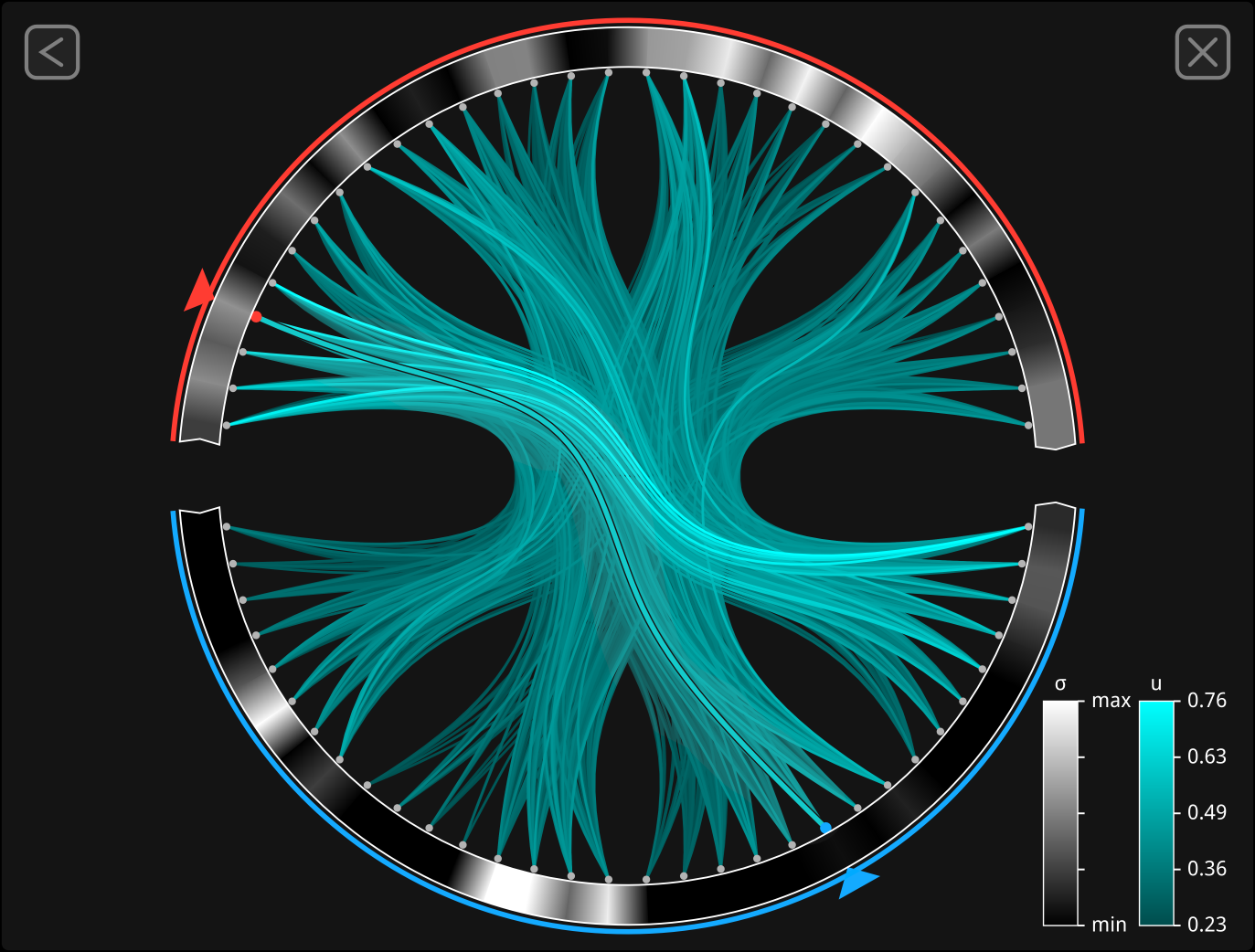}};
  \node [inner sep=0pt,right=of a] (b) {\includegraphics[height=3.6cm]{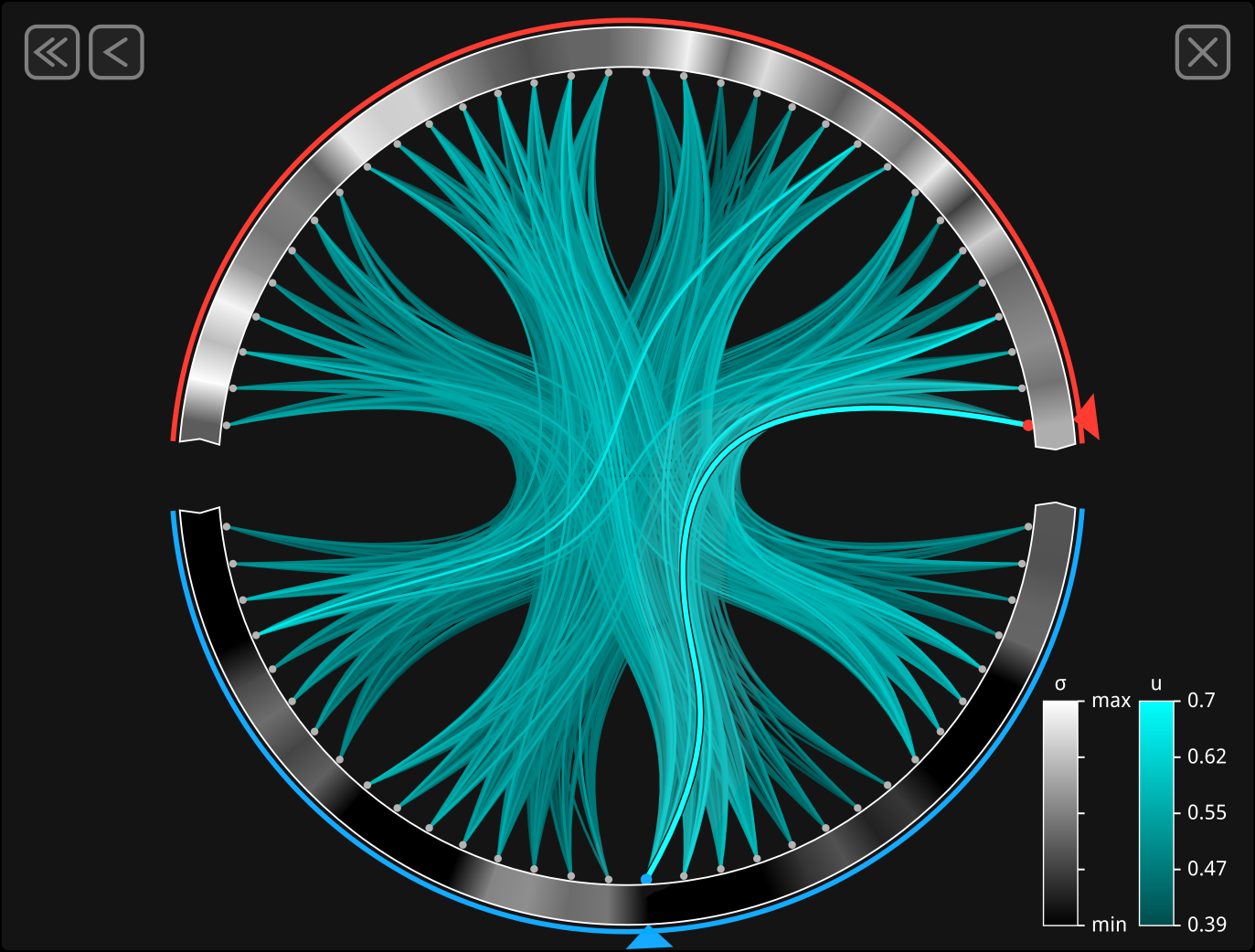}};
  \node [inner sep=0pt,below=of b] (c) {\includegraphics[height=3.6cm]{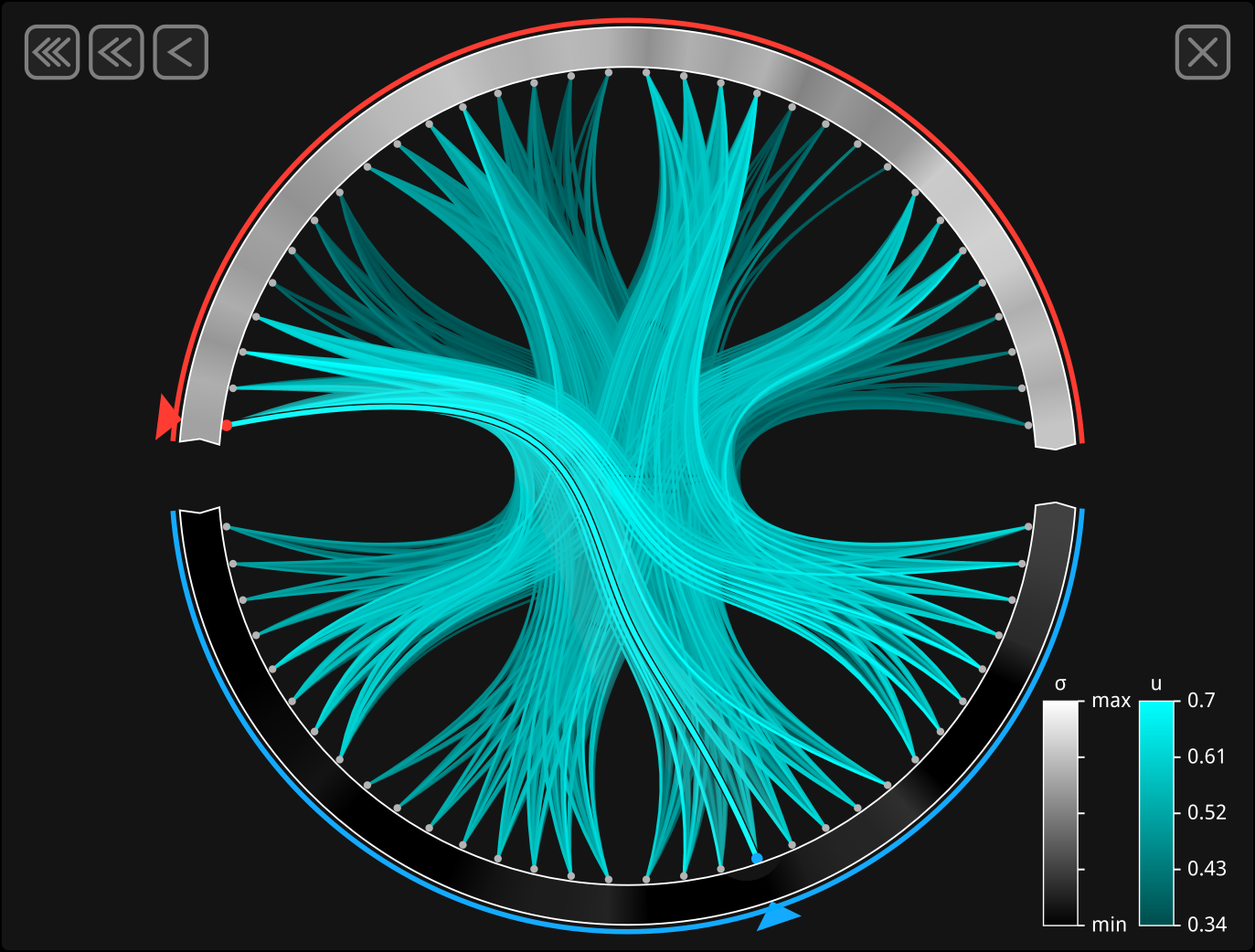}};
  \node [inner sep=0pt,left=of c] (d) {\includegraphics[height=3.6cm]{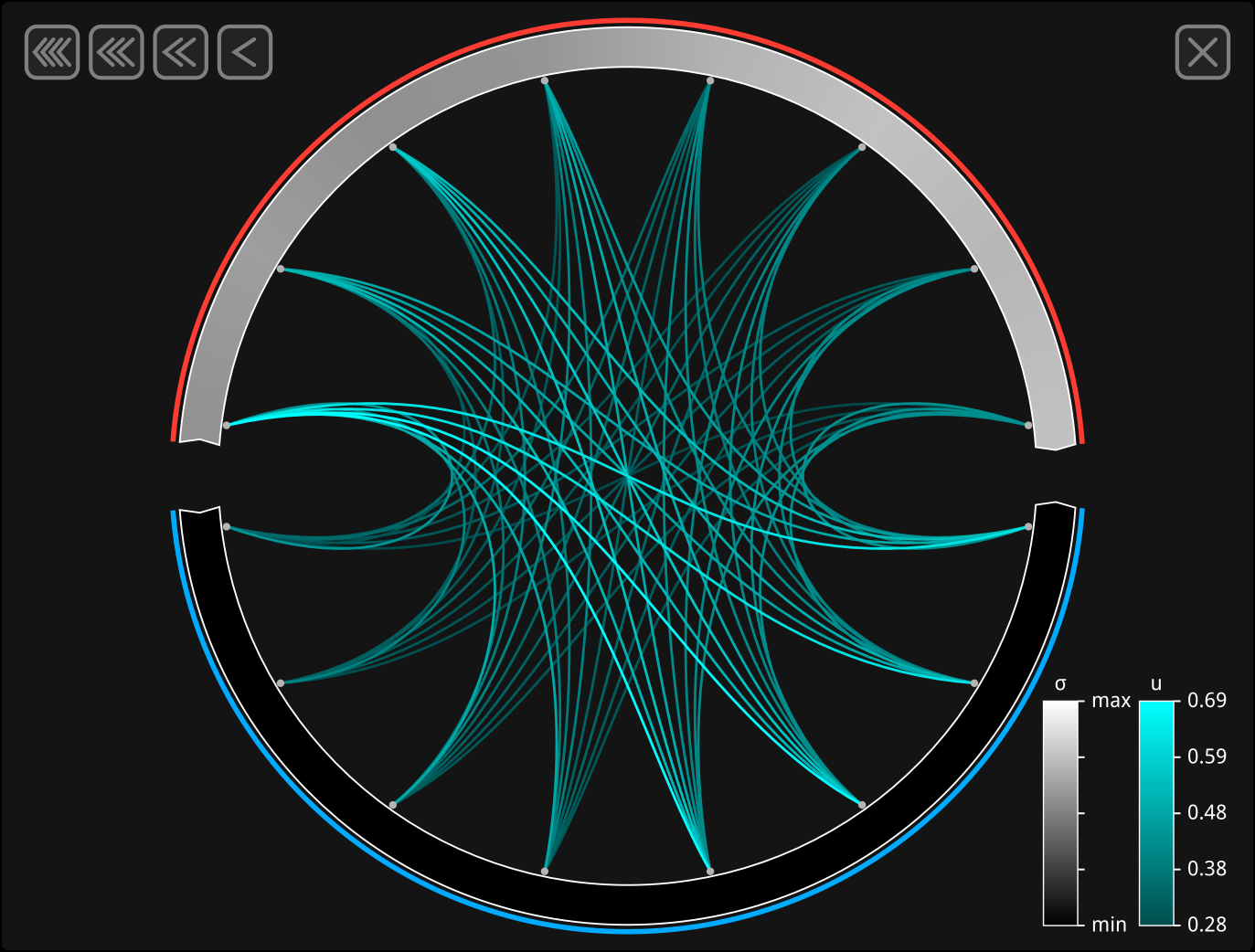}};
\end{scope}
\draw[->,cyan,ultra thick] (a.east) -- (b.west);
\draw[->,cyan,ultra thick] (b.south) -- (c.north);
\draw[->,cyan,ultra thick] (c.west) -- (d.east);
\node[fit=(a)(b)(c)(d)](group){};

\node [left=of group,inner sep=0pt] (z) {\includegraphics[height=7.3cm]{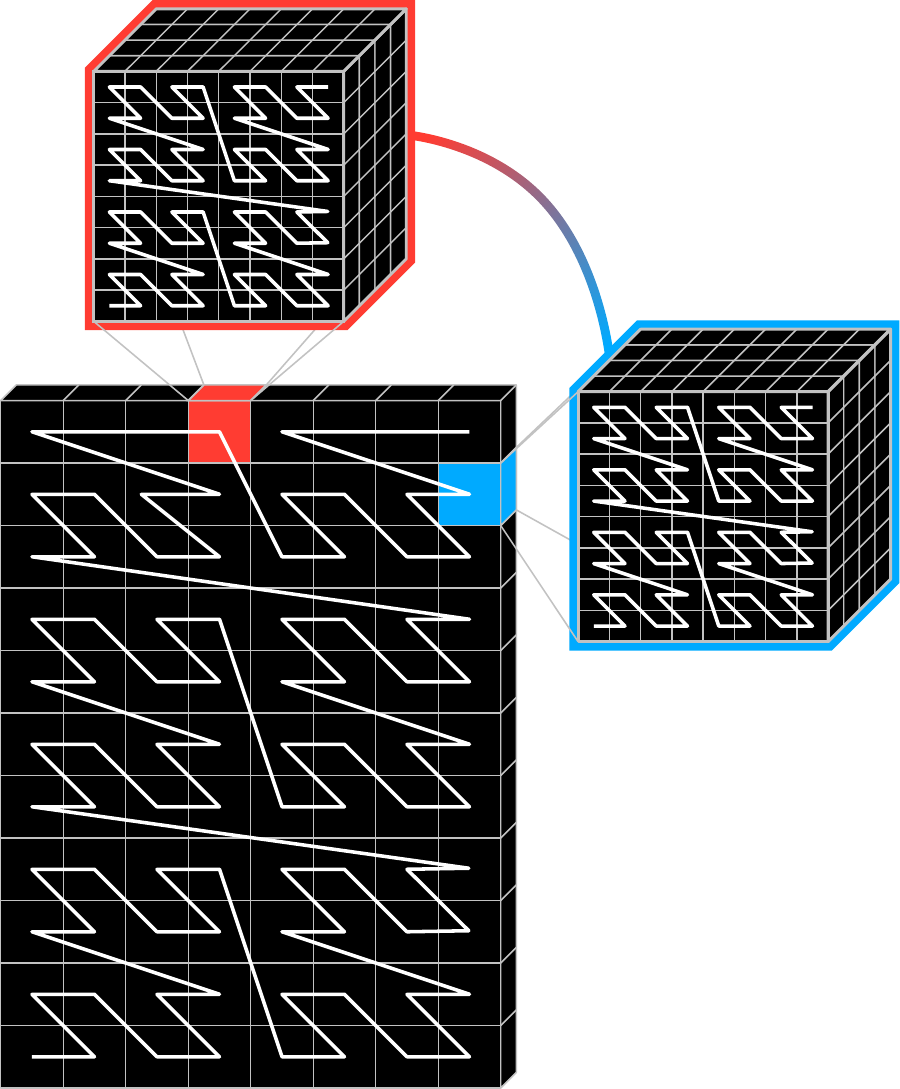}};

\gettikzxy{(a.west)}{\ax}{\ay}
\gettikzxy{(z.east)}{\zx}{\zy}
\draw[->,black,ultra thick] (\zx,\ay) -- (a.west);


\end{tikzpicture}
\caption{Recursive linear layout of bricks along a z-curve. Two bricks are selected (red and blue), and laid out on the respective upper and lower half-circle in the top left  focus view. Entities in each brick are linearized via a z-curve. In further refinements of the focus view (indicated by blue arrows), the picked regions are again laid out along z-curves. 
}
\label{fig:zorder-refinement}
\end{figure*}


\subsection{Focus Chord Diagram}

When selecting an edge in the context chord, the correlations between the entities in the two bricks connected by this edge (called bricks $A$ and $B$ in the following) are displayed in finer detail in a new chord diagram---the focus chord diagram. While the context diagram displays the correlations between all $M(M-1)/2$ pairs of $M$ initial bricks, the focus diagram displays the $M/2$ refined bricks of $A$ on the bottom semicircle and the $M/2$ refined bricks of $B$ on the top semicircle of the circular chord layout. Self-correlations within $A$ and $B$ are not considered, as selecting an edge indicates a user's interest in a finer representation of the correlations between the two connected regions, but not necessarily the self-correlation within these regions. The selection triggers the refinement of $A$ and $B$ into $M$ smaller bricks and $M/2 \times M/2$ pairs for which correlations are computed and edges are drawn.
For computing the correlations, correlation sampling is used as described in \cref{sec:bayopt-strat}.
The ring is again colored via the ensemble spread in each of the smaller bricks.

As shown in \cref{fig:zorder-refinement}~right, by first selecting an edge in the context view and then subsequently selecting edges in the focus view, the user can adaptively navigate down to the finest resolution level.
The focus chords, however, are not shown side by side, but each newly refined chord replaces the currently seen one. To show the refinement level, as many navigation buttons (visualized via arrow symbols) are shown as the number of times the user has selected a refinement. By clicking 
a buttons, the user can jump back in the hierarchy by the number of levels that is encoded into the arrow symbols.

Each selected brick is laid out along one semicircle using a local Z-curve, as shown in ~\cref{fig:zorder-refinement} (left). To indicate which bricks have been selected for refinement in the context view, 
two arrows in blue and red are used to mark these bricks. 
Accordingly, the top and bottom semicircle in the focus view are visualized using a blue or red outline (cf.~\cref{fig:zorder-refinement} (right)). The spatial regions corresponding to the selected bricks are put into spatial context in a 3D view, by drawing them as opaque boxes with the size and at the locations of the selected bricks. 
The 3D view is updated whenever the user either selects an edge or hovers over an edge with the mouse cursor in the context or focus view.

In case the user is interested in refining only one single brick, i.e., to specifically look at self-correlations in this brick, the small circle representing the brick in the context or focus diagram can be clicked. In this case, the new focus diagram is generated as previously described, yet now the refined bricks of the selected one are shown along the top and bottom semicircle in the focus view. 

\subsection{Comparative Correlation Visualization}


The proposed workflow for correlation sampling and visualization can be used directly to compare spatial correlations in two different fields, i.e., two different simulation variables. In this case, each brick is duplicated to also represent the second variable field,  
and the edges of the different variables are rendered into the same chord diagram with different colors.

\begin{figure}[t]
\centering
\includegraphics[height=3.5cm]{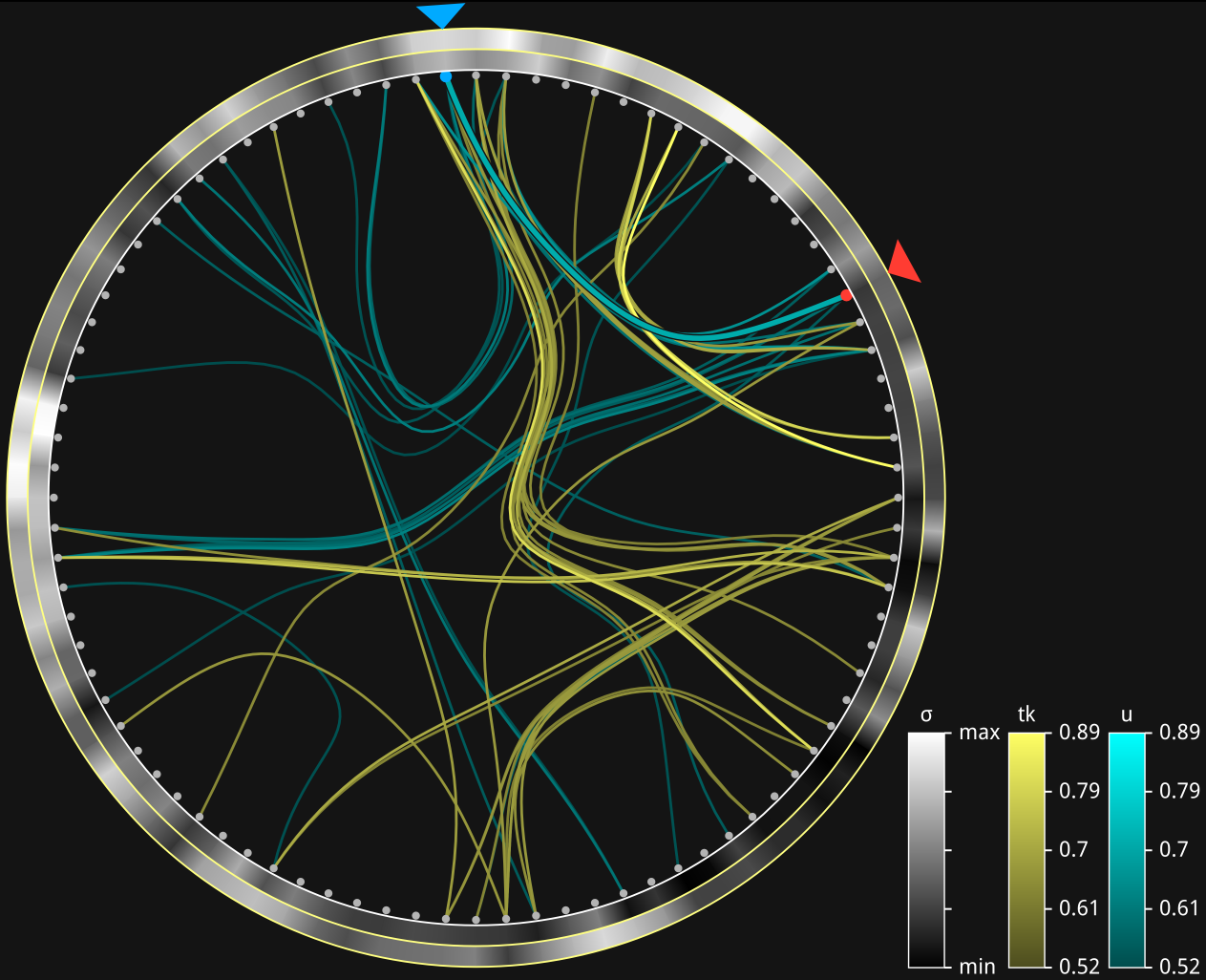}\hspace{0.04cm}
\includegraphics[height=3.5cm]{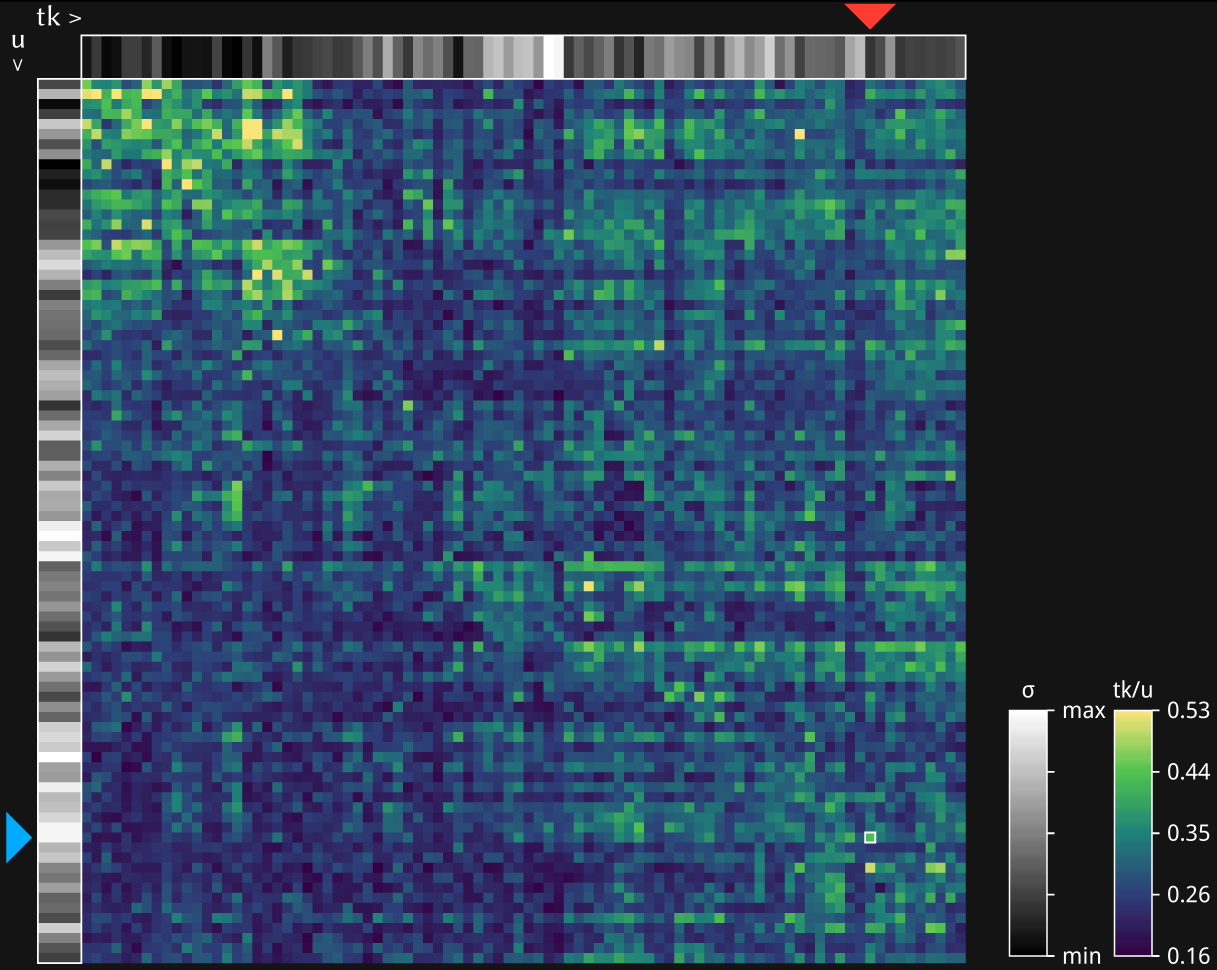}
\caption{Left: Comparative visualization of MI in temperature \texttt{tk} and longitudinal wind component \texttt{u} of a weather forecast ensemble~\cite{Necker2020}. Right: Inter-variable correlation matrix showing correlations between \texttt{tk} and \texttt{u}. Entries in the upper right triangular structure indicate correlations between \texttt{tk} at first region (column) to \texttt{u} at second region (row), and between \texttt{u} at first region (row) to \texttt{tk} at second region (column) for the lower left triangular structure.}
\label{fig:comparative}
\end{figure}

Since edges showing different variables can occlude each other they are sorted with respect to (absolute) correlation strength and rendered in the order of increasing strength. Thus, edges indicating higher correlation are blended on top of curves with lower correlation. This is demonstrated in \cref{fig:comparative}~left. Furthermore, a second ring is shown in the chord diagram to indicate the ensemble spread with respect to a second variable, enabling an effective comparison of the spatial distribution of the spread. 

In addition to showing edges that indicate correlations in a single variable field, it is interesting to analyze the correlations between two different variables. BOS (or sampling on aggregates) is then performed as described, but point-to-point correlations are computed between pairs of points from either field. When a chord diagram is used to visualize these correlations, however, the same edge is displayed with different correlation strengths, i.e., to indicate the correlation between the first and the second variable at, respectively, the first and the second point, as well as the correlation between the first and the second variable at, respectively, the second and the first point. To avoid this, we show variable-to-variable correlations in a correlation matrix instead of a chord diagram. This matrix, which we subsequently call inter-variable correlation matrix, is completely filled due to the mentioned anti-symmetry and does not show redundant information (see \cref{fig:comparative}~right). 
Once the system switches from a chord diagram to an inter-variable correlation matrix, adaptive refinement is supported as described. By selecting a matrix element, the refined regions are shown in a focus diagram using again a correlation matrix.

\section{Analysis and Results}

\subsection{Data Sets}\label{sec:datasets}

We analyse the proposed correlation visualization using four different data sets: 
``Synth1'' is a synthetic 1000 member ensemble at a grid resolution of $256 \times 256 \times 32$. It is used to test the ability of BOS to find the true maxima between brick pairs. It contains two larger clusters and close-by smaller mini-clusters with high positive correlation (cf.~\cref{fig:synthetic-4x4}~bottom). Within each cluster, the correlations are decaying by their $l_\infty$-norm distance from the cluster center (i.e., $\max\{d_x, d_y, d_z\}$). The mutual correlation between each pair of clusters is high as well. Between clusters there are regions with low correlations to any of the clusters. ``Synth1'' has been generated by computing at each grid point one sequence of 1000 uncorrelated random samples, another sequence of 1000 perfectly linearly correlated deterministic samples, and a number $\lambda \in [0, 1]$ that is used to interpolate between the values in these sequences ($\lambda=1$ at cluster centers, decaying by the distance to the closest center). Using deterministic, identical samples in the case of $\lambda=1$ makes sure that the clusters are mutually correlated.

``Synth2'' is a second synthetic data set that is used to examine the convergence properties of BOS (cf.~\cref{fig:sampling}). Computing the ground truth maximum between two bricks of size $n^3$ has a time complexity of $O(n^6)$, which quickly becomes infeasible. Thus, we generate ``Synth2'' so that the maxima for each pair of bricks can be computed analytically. In particular, we have chosen the probability density function of a 6-variate normal distribution as a substitute for correlation computation. This function is cheap to evaluate, the density maximum is known to be at the distribution mean, and the minimum is located at one of the domain corners. We don't use mixtures of Gaussians, as finding the maxima in such distributions is significantly more complicated \cite{ModeFindingGaussianMixture1, ModeFindingGaussianMixture2}. The mean and covariance of the distributions are sampled randomly during our experiments.

``Necker''~\cite{Necker2020} is a convective-scale 1000 ensemble member simulation forecast over central Europe, using a ``full-physics non-hydrostatic regional model (SCALE-RM) with a horizontal grid spacing of 3 km''. The values are stored on a regular grid of size $250 \times 352$ with 20 vertical height levels that capture vertical variations. The ensemble comprises multiple physical variables, such as temperature, wind speed, hail and precipitation. ``Matsunobu''~\cite{MatsunobuEnsemble} is a large ensemble that has been simulated using the ICON-D2 numerical weather prediction model. It stores 180 ensemble members at a grid of size $640 \times 704$ with 64 vertical height levels. 
While ``Necker''---even when two variables are compared---fits entirely into the memory of our target GPU, this doesn't hold for ``Matsunobu''. Thus, the mean hierarchy described in \cref{sec:aggregate-sampling} is used in this case to generate the context view.

\subsection{Performance Evaluation}\label{sec:perf}

For PPMCC and KMI, \Cref{tab:perf} shows the performance that is achieved on the CPU and the GPU. All timings have been evaluated on a system running Ubuntu 20.04 with an AMD Ryzen 9 3900X 12-core (24-thread) CPU and an NVIDIA GeForce RTX 3090 GPU. The experiments have been performed with ``Necker'', by selecting a grid point in the 3D domain and computing the correlations to all other grid points. This amounts to 1.76 million correlation computations, using 100 and 1000 members, respectively. 


For computing KMI on the CPU, a parallelization is used that allocates dynamic heap memory outside of the parallelization loop. This avoids expensive system calls which affect the performance in every loop iteration. All CPU implementations utilize OpenMP for parallelization. The GPU implementation of KMI shows a considerable speedup of approximately $11 \times$.

For the 1000 member ensemble, the throughput even for the computationally most expensive measure KMI is about
$49000$
samples per second. Notably, however, this throughput cannot be achieved with any of the tested sampling methods. The reason is that the one-to-many query is extremely cache-friendly and can effectively exploit SIMT parallelism on the GPU. Furthermore, when used in combination with chord diagrams, computed correlations need to be downloaded to the CPU for generating the diagrams. In practice, a throughput between roughly 18000 samples per second is achieved for random sampling. BOS achieves a throughput of 11000 samples per second for BOS at 100 samples per brick pair and 8300 at 400 samples per brick pair. When using BOS, it is in particular the computational overhead that is introduced by the evaluation of a Gaussian process model that leads to the reduction, which scales non-linearly with the number of samples added to the model.

\begin{table}[t]
\caption{Performance comparison of CPU and GPU implementations of PPMCC and KMI. All correlations from a single reference point to all other 1,760,000 grid points in ``Necker'' are computed. $n$ is the number of ensemble members.}
\label{tab:perf}
\begin{tabularx}{\linewidth}{X|X|X|X|X}
& Measure & CPU (s) & GPU (s) & Speedup \\
\hhline{=|=|=|=|=}
\multirow{2}{*}{$n=100$}  & PPMCC & 0.049 & 0.002 & 24.5x \\
                          & KMI   & 6.085 & 0.672 & 9.1x \\
\hline
\multirow{2}{*}{$n=1000$} & PPMCC & 2.1 & 0.21 & 10.0x \\
                          & KMI   & 412.2 & 35.9 & 11.5x
\end{tabularx}
\end{table}
The situation is slightly different for the data set ``Matsunobu'', which cannot be stored entirely in GPU memory. 
Thus, the system estimates correlations from the level where $2 \times 2 \times 2$ data values are represented by their means. Then, the entire data set fits into GPU memory and the context diagram can be computed in approximately 15 seconds. 

\subsection{Qualitative Evaluation}\label{sec:bayopt-eval}







\begin{figure*}[t]
\centering
\includegraphics[height=4.85cm]{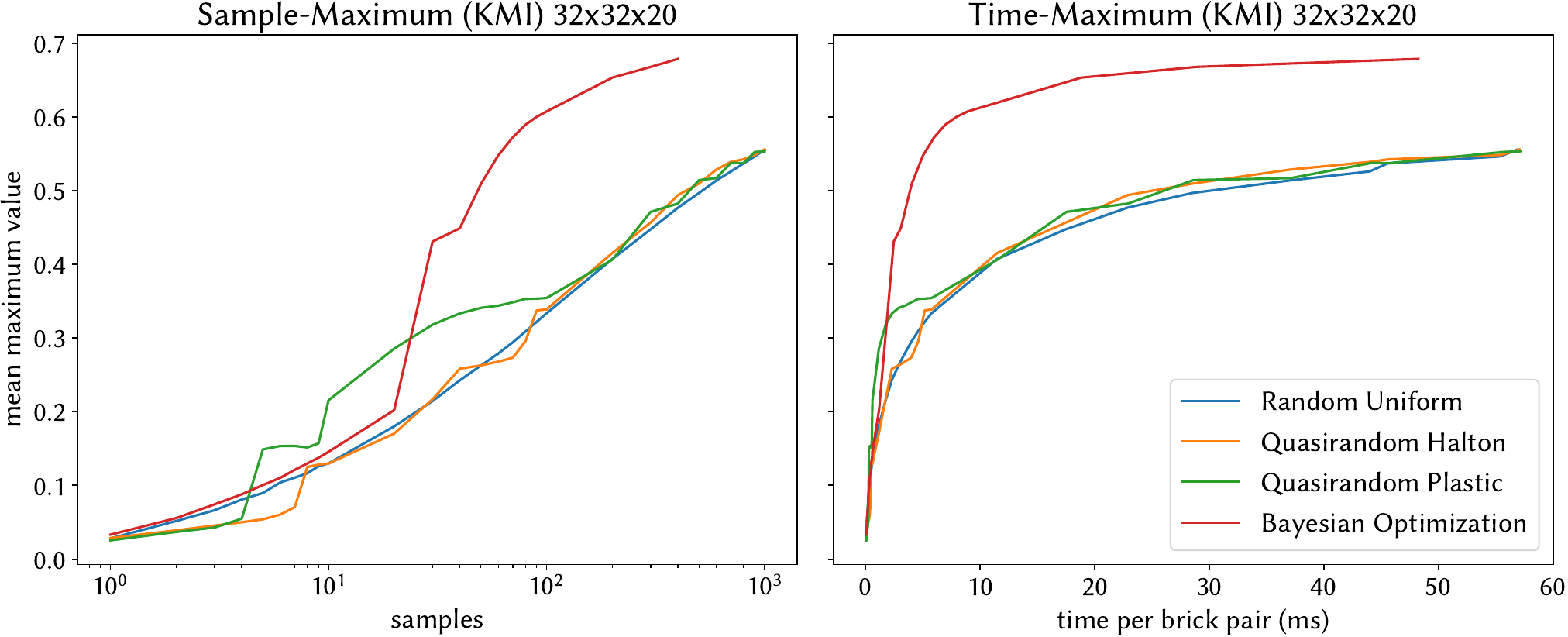}
\hspace{0.1cm}
\includegraphics[height=4.85cm]{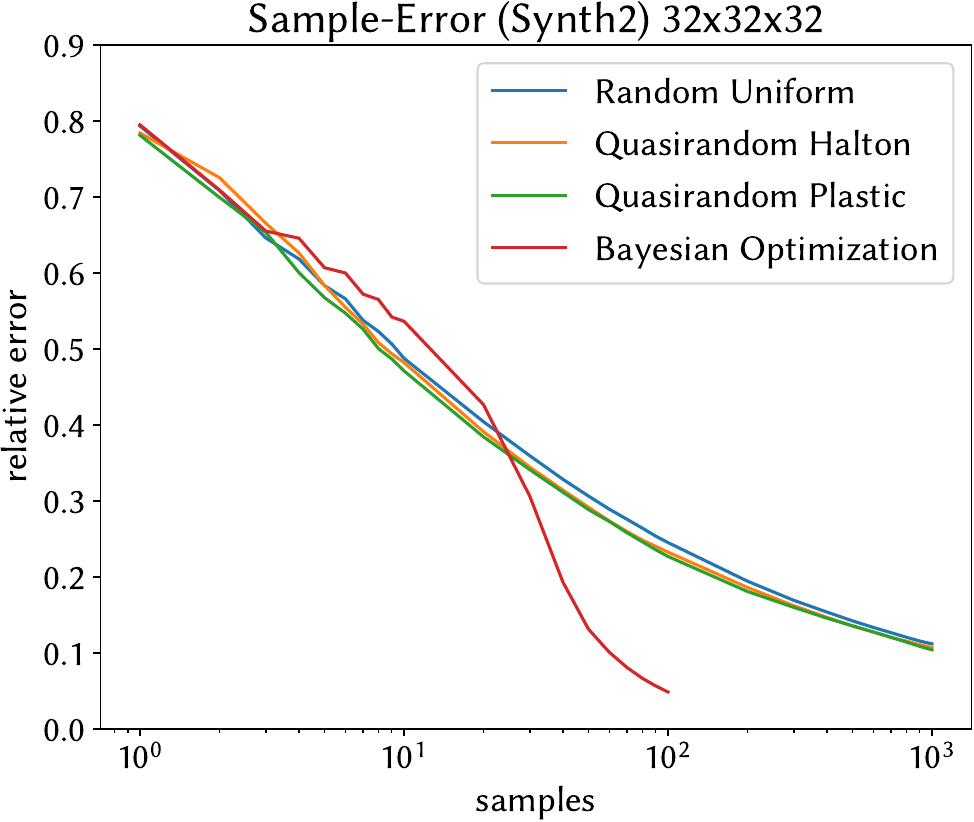}
\caption{Left and middle: Correlation sampling via BOS and random sampling variants. Mean sampled maximum for all brick pairs of $32 \times 32 \times 20$ voxels in ``Necker''. Higher maximum is better.
All observations are averaged over 10 runs. For the same number of samples, BOS achieves higher maxima in a shorter amount of time. Right: Mean normalized error for 1000 randomly selected brick pairs of $32 \times 32 \times 32$ voxels in ``Synth2''. Lower error is better. After an initial lead of random sampling, BOS converges to the almost exact maximum with only 100 samples.
}
\label{fig:sampling}
\end{figure*}

Firstly, we evaluate the sampling strategies introduced in Sec.~\ref{sec:bayesian} for computing representative correlation maxima between pairs of bricks. 
For comparing the quality of different sampling strategies, it is necessary to compute the ground truth beforehand, i.e., the minimum and maximum of all point-to-point correlations between any pair of grid points in two bricks. Then, the relative error of the sampled maxima can be computed as the fraction of the difference between the maximum and minimum. However, for bricks of size $32 \times 32 \times 20$, as, for instance, used to generate the context view of ``Necker'', this already amounts to more than 400 million point-to-point correlations that need to be computed for a single pair of bricks. Since in the statistical evaluation we consider averages over many pairs of bricks, this is unfeasible due to time constraints. On the other hand, considerably smaller brick sizes, for which the ground truth can be computed in a reasonable amount of time, are no longer representative. Consequentially, for evaluation with a real data set as shown in \cref{fig:sampling}~left and middle, we show the convergence rates with increasing number of samples and increasing time budget that are achieved with different sampling strategies. The ground truth to which the sampling schemes converge is not known. In these experiments, the Kraskov MI estimator is used with ``Necker''.

As can be seen in \cref{fig:sampling}~left, for an equal amount of samples, BOS comes significantly closer to the assumed maximum than all alternative sampling variants. The resulting curve is steeper and indicates faster convergence to the ground truth than those of the alternatives. To confirm that the lower number of samples required by BOS is not outweighed by higher computation times for each sample, \cref{fig:sampling}~middle shows the estimate of the maximum for given time budgets. Notably, BOS achieves higher accuracy than all other variants in the same amount of time, despite the fact that BOS samples are considerably more expensive as shown in the performance analysis before.

We use ``Synth2''  to further assess the convergence properties of all sampling methods.
As shown in \cref{fig:sampling}~right, BOS converges faster than all other variants. 
Here, the y-axis shows the mean relative error of the computed maxima. For each approximated maximum obtained via sampling, we compute the normalized relative distance on the scale from the exact minimum and maximum for the specific brick pair. A value close to 1 indicates that the sampling method is close to the minimum on average, and a value of 0 that it is close to the maximum. It can be seen that, on average, random sampling gives slightly better results than BOS for a low number of samples. BOS, on the other hand, shows the trade-off between exploration and exploitation. Exploration means searching in previously relatively unexplored areas with high uncertainty for possibly higher maxima. Exploitation means refining an already known good maximum to find a possibly higher value in its surroundings. For the synthetic data set, the acquisition function used by the Bayesian optimizer might prioritize exploration in the beginning. Since the data set is smooth and has a single local maximum, this only seems to pay off for a higher number of samples.

\begin{figure}[t]
\centering
\includegraphics[width=0.5\columnwidth]{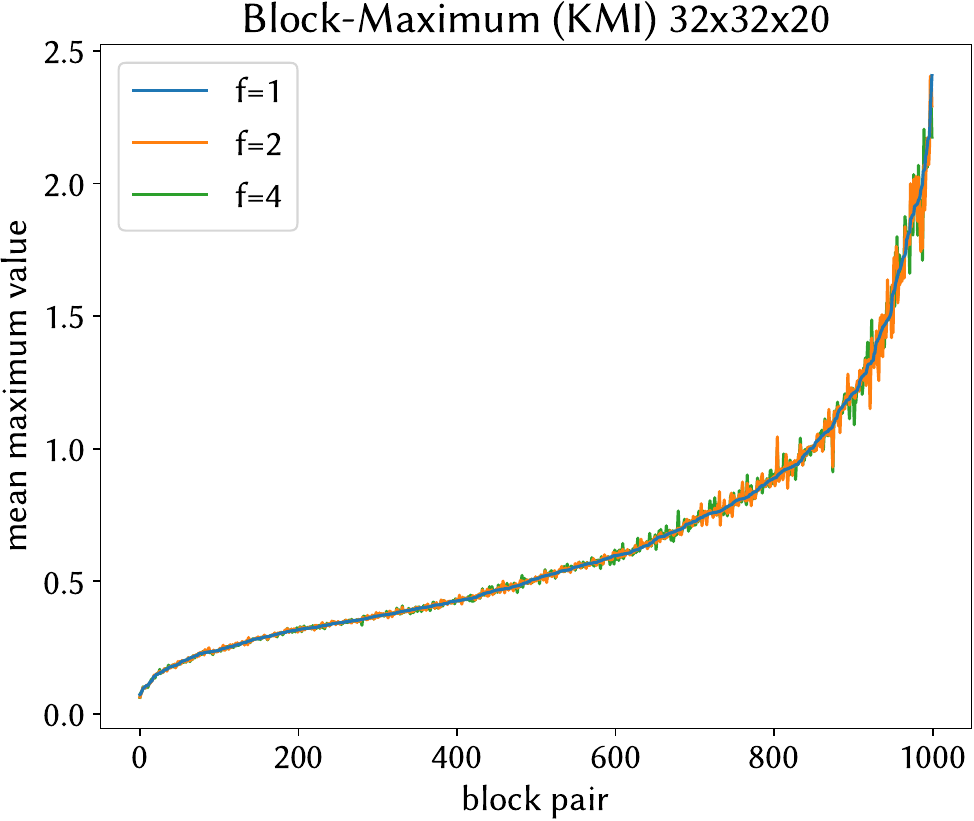}
\caption{Sampled correlation maxima in 1000 brick pairs from ``Necker'' (averaged over 10 runs) using the initial data ($f=1$) and means of $2 \times 2 \times 2$ ($f=2$) and $4 \times 4 \times 4$ ($f=4$) data values in the temperature field \texttt{tk}. Brick pairs are sorted wrt increasing maxima in the initial data to reduce spikes. The average deviation for $f=2$ and $f=4$, respectively, is $1.6\%$ and $1.7\%$.}
\label{fig:mean-corr}
\end{figure}

Furthermore, we evaluate the accuracy of correlation sampling when brick-to-brick correlations are computed via BOS on mean values instead of the initial data values. The maximum correlation between 1000 randomly sampled brick pairs of size $32 \times 32 \times 20$ in ``Necker'' is computed using BOS with 100 samples. BOS is performed on the initial data values, and on brick pairs of size $16 \times 16 \times 10$ and $8 \times 8 \times 5$ where respectively each entry contains the mean of $2 \times 2 \times 2$ and $4 \times 4 \times 4$ initial data values. All brick-to-brick correlations are computed 10 times and then averaged to reduce the effect of random initializations of BOS. 
The brick pairs are then sorted with respect to increasing maxima when computed on the initial data, and the maxima that have been computed on mean values are arranged accordingly. The results are shown in \cref{fig:mean-corr}, demonstrating that even at a downsampling factor of four the maxima that are computed on the corresponding mean values are in good agreement with the maxima computed on the initial data. At the same time, even at a downsampling factor of two the memory requirement is reduced by a factor of eight, so that ``Matsunobu'' can be stored entirely in GPU memory.


\subsection{Sampling Strategy}\label{sec:bayopt-strat}
Our tests confirm that even though BOS samples are more expensive than samples taken via random sampling, BOS requires so much less samples that overall the same quality is achieved in less time. However, when repeating the experiments with ever smaller bricks, the performance gains of BOS become less and less significant. In particular, while a noticeable advantage of BOS can still be perceived at a brick size of $16 \times 16 \times 16$, beyond that the random sampling approaches become more efficient at equal quality. This is because the data values in smaller bricks tend to have lower variation, in general, and the overall number of required samples becomes so low that the computational overhead of BOS cannot be amortized. 

According to this observation, we pursue the following strategy. BOS is used to estimate the correlation maxima between the many pairs of large bricks at the context level. For instance, at the context level of ``Necker'' there are $8 \times 11$ bricks of size $32 \times 32 \times 20$, so that 88x87/2 brick-to-brick correlation maxima need to be estimated. With BOS using 100 samples per brick pair and KMI as correlation measure, this requires roughly 16 seconds of pre-processing time. Notably, it would take about 365 days to compute the brick-to-brick maxima if all point-to-point correlations were to be considered. At focus levels, and at latest if the brick size goes below $16 \times 16 \times 16$, the system switches to random sampling down to the level where the computation of correlations between all point-to-point pairs is fast enough. Here we consider an update time of less than 5 seconds to be acceptable.

\begin{figure}[t]
\centering
\includegraphics[width=0.32\linewidth]{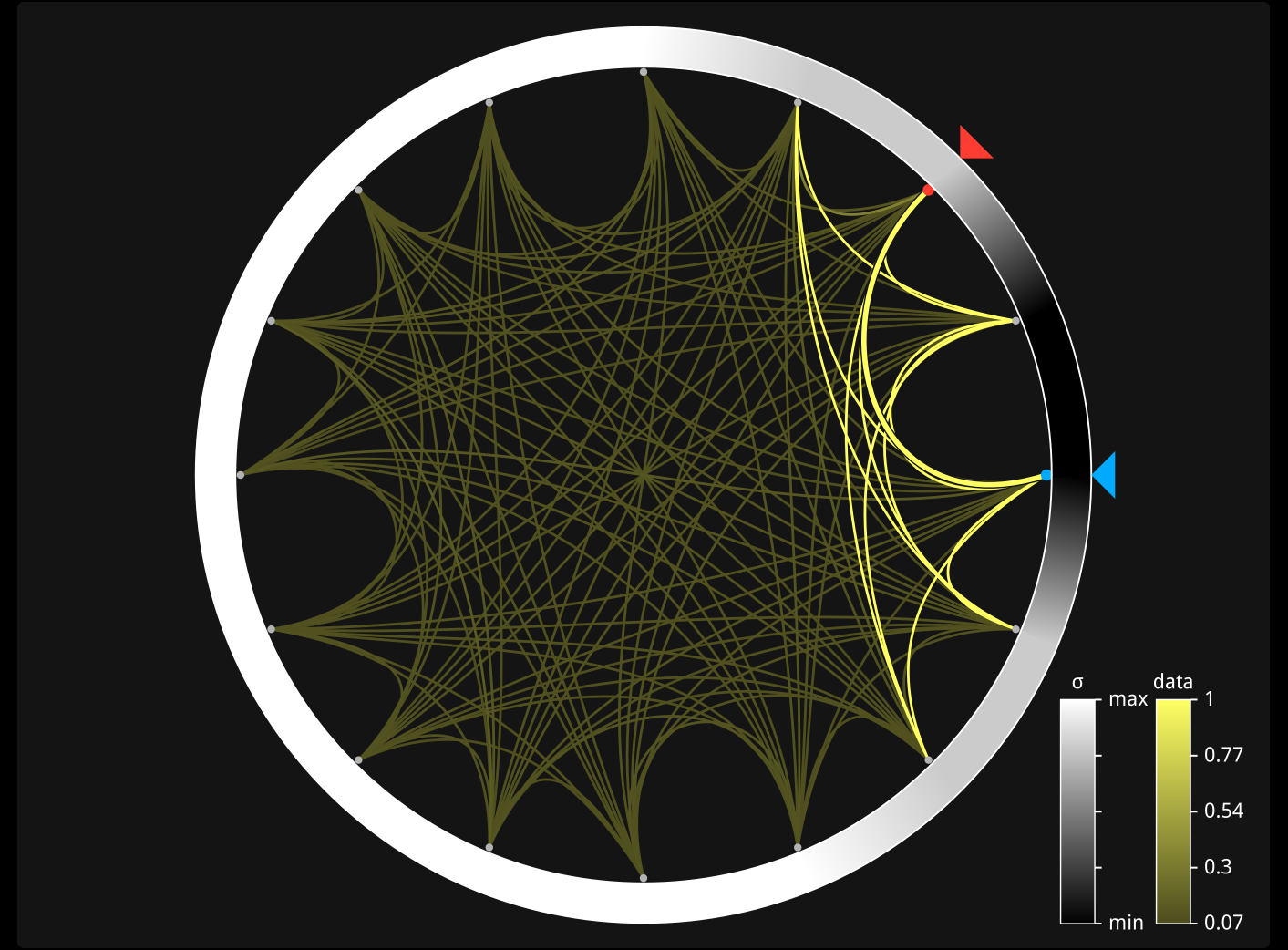}
\includegraphics[width=0.32\linewidth]{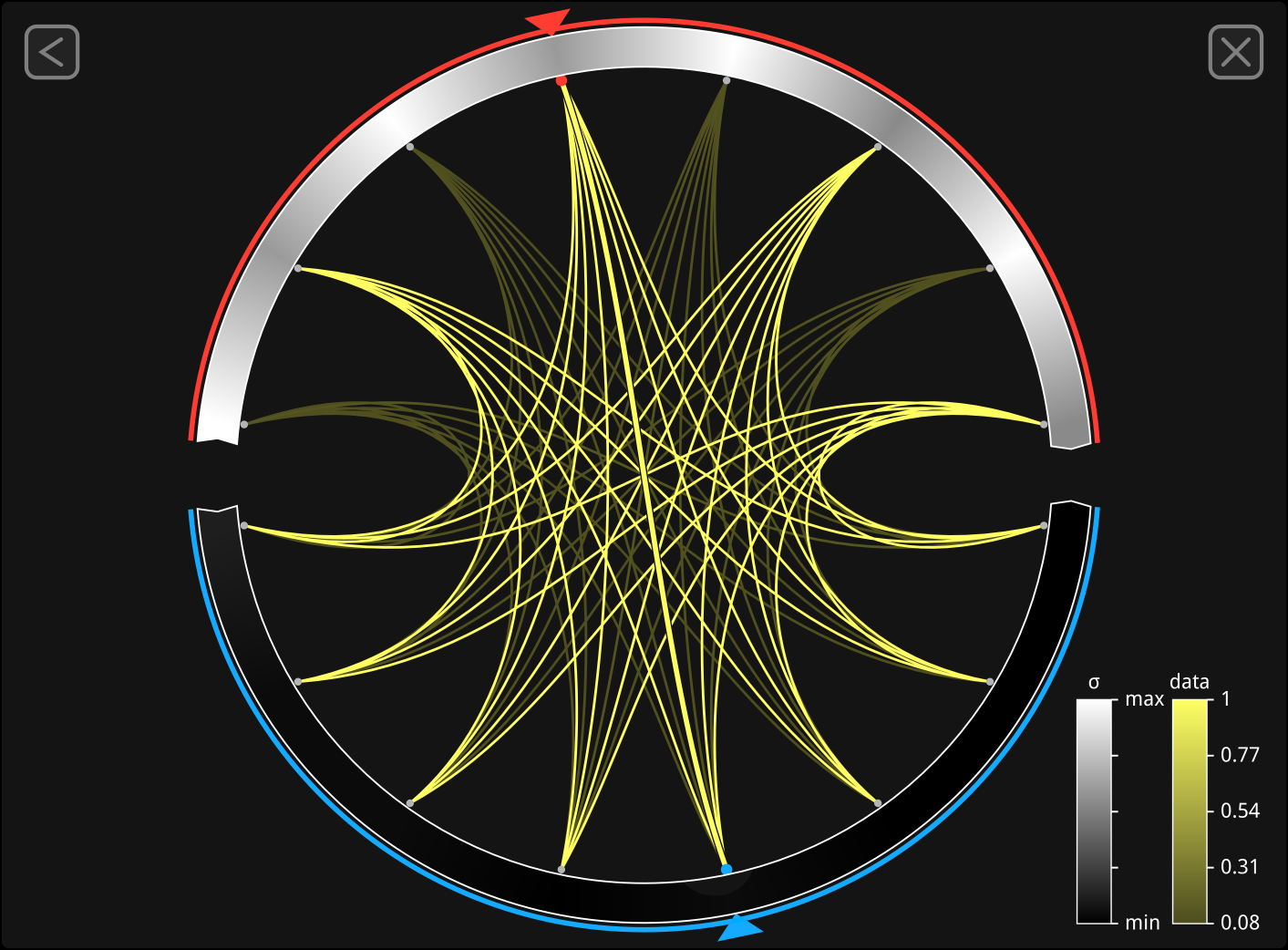}
\includegraphics[width=0.32\linewidth]{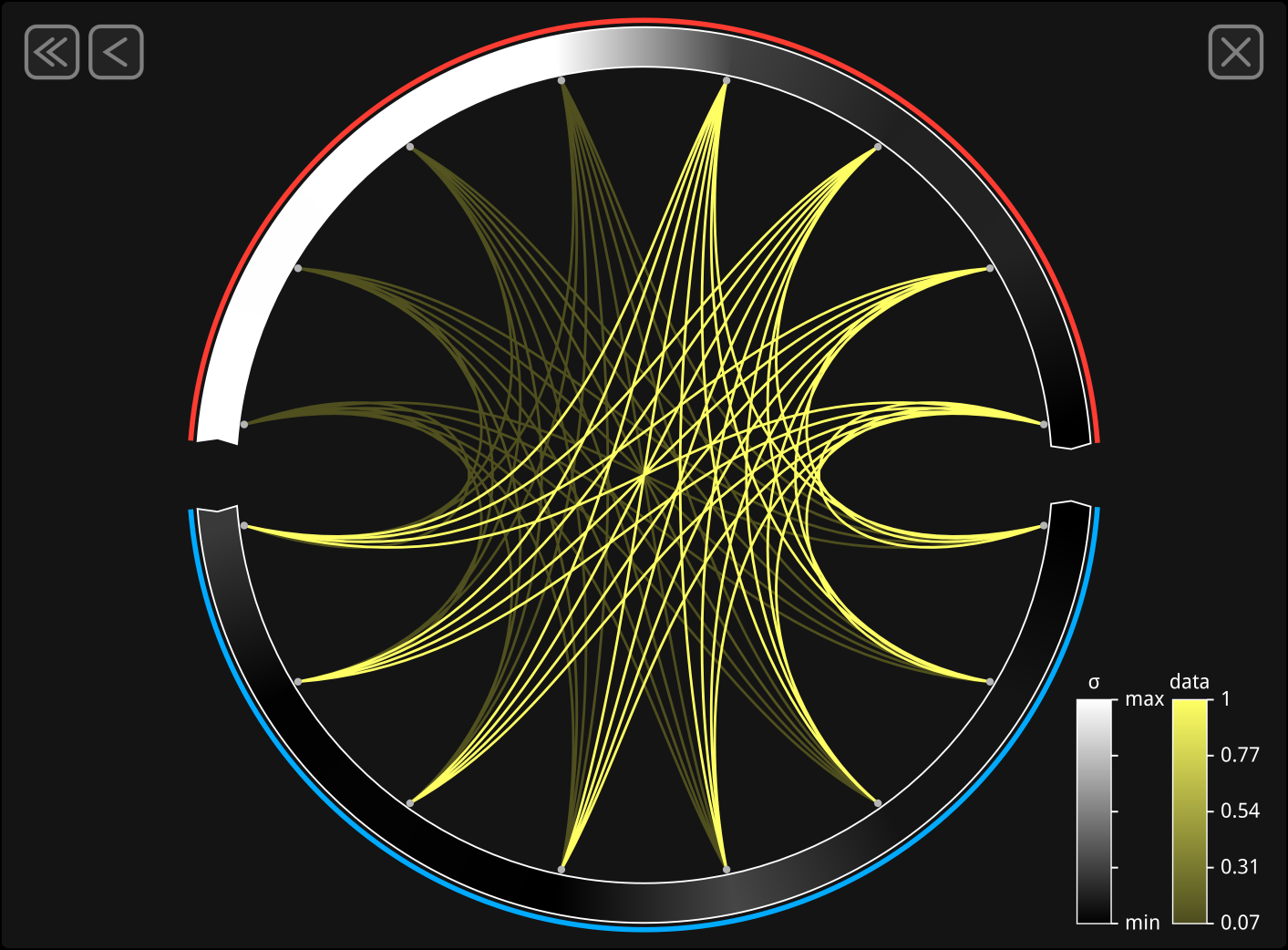}\vspace{0.08cm}
\includegraphics[width=0.32\linewidth]{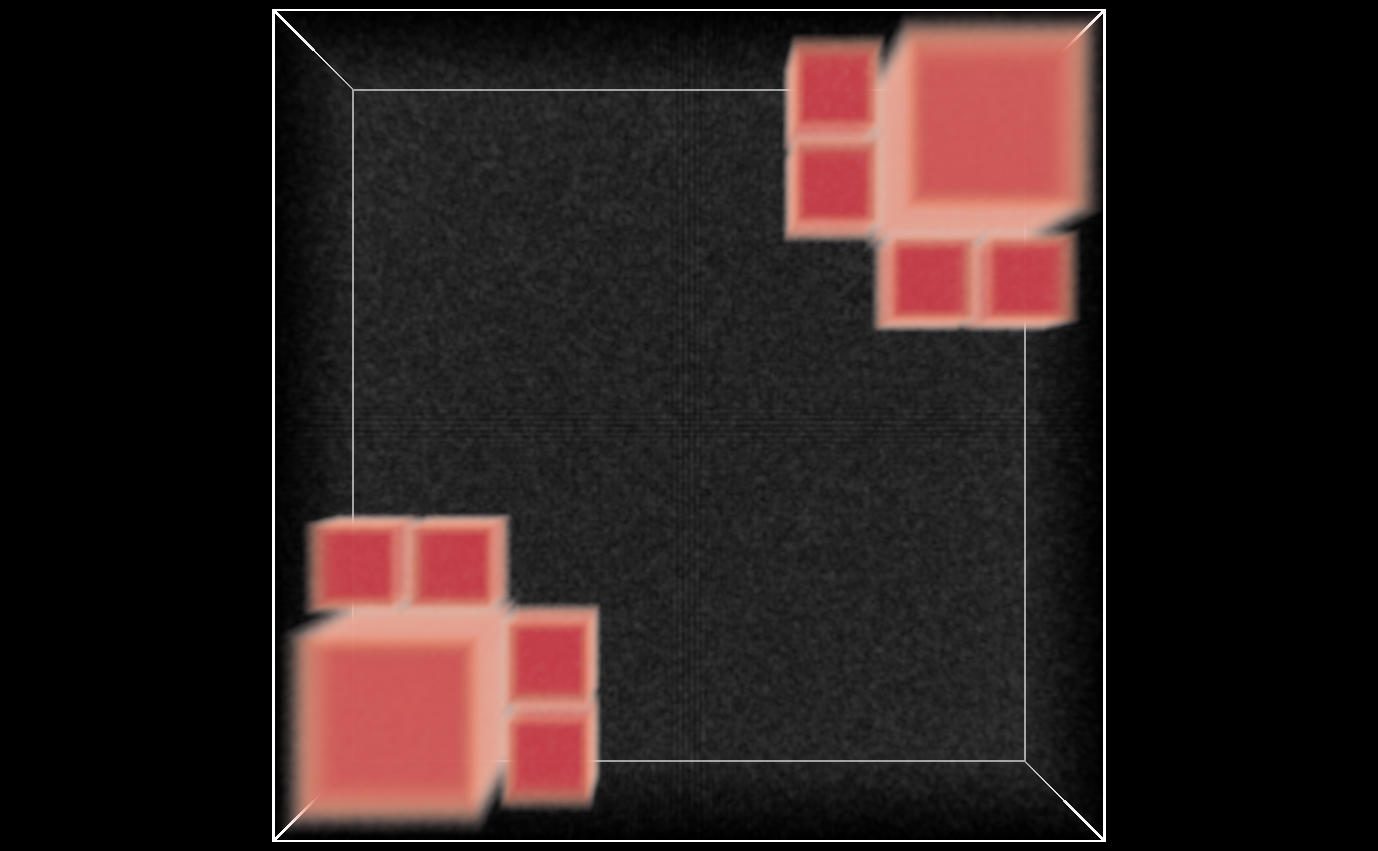}
\includegraphics[width=0.32\linewidth]{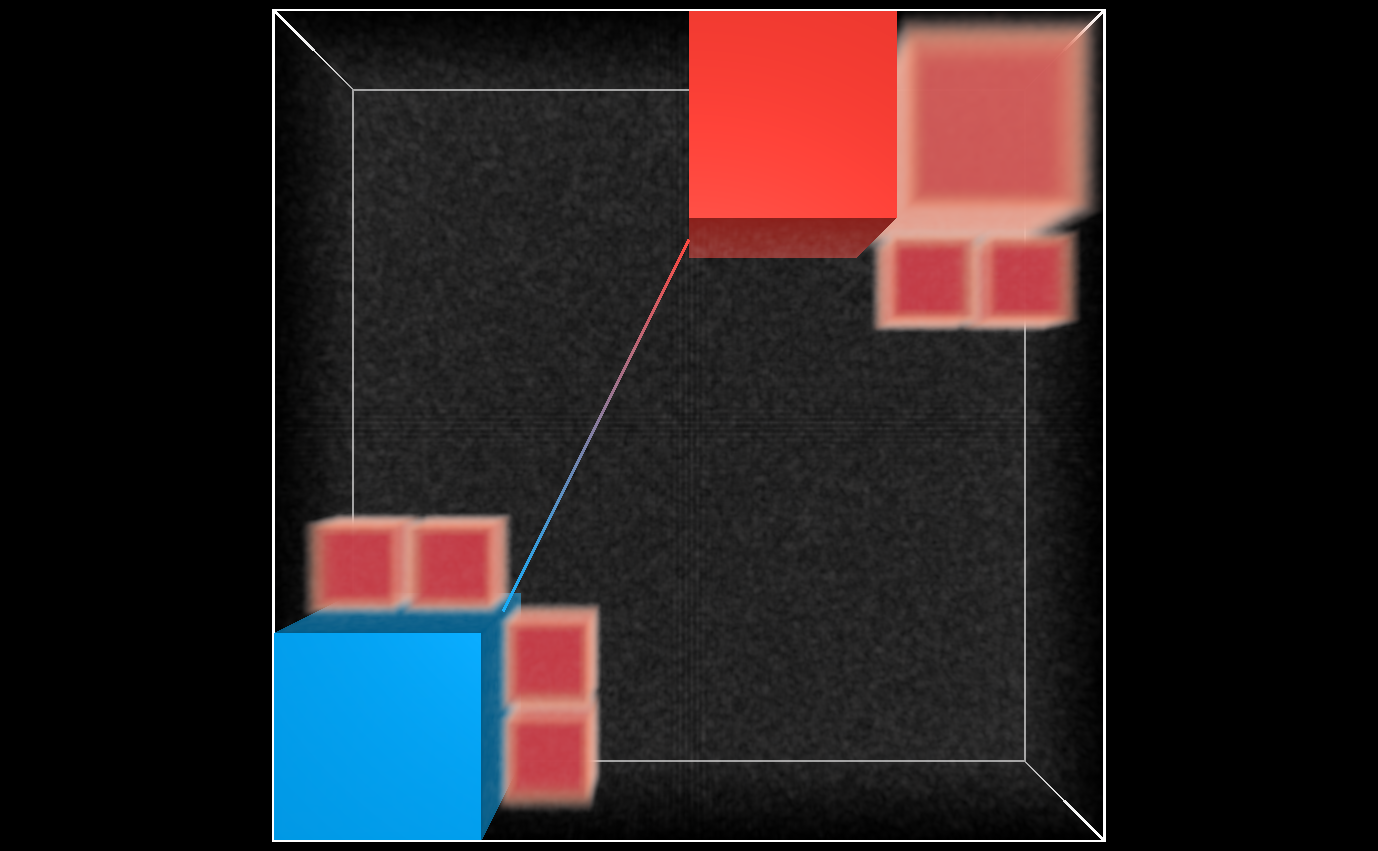}
\includegraphics[width=0.32\linewidth]{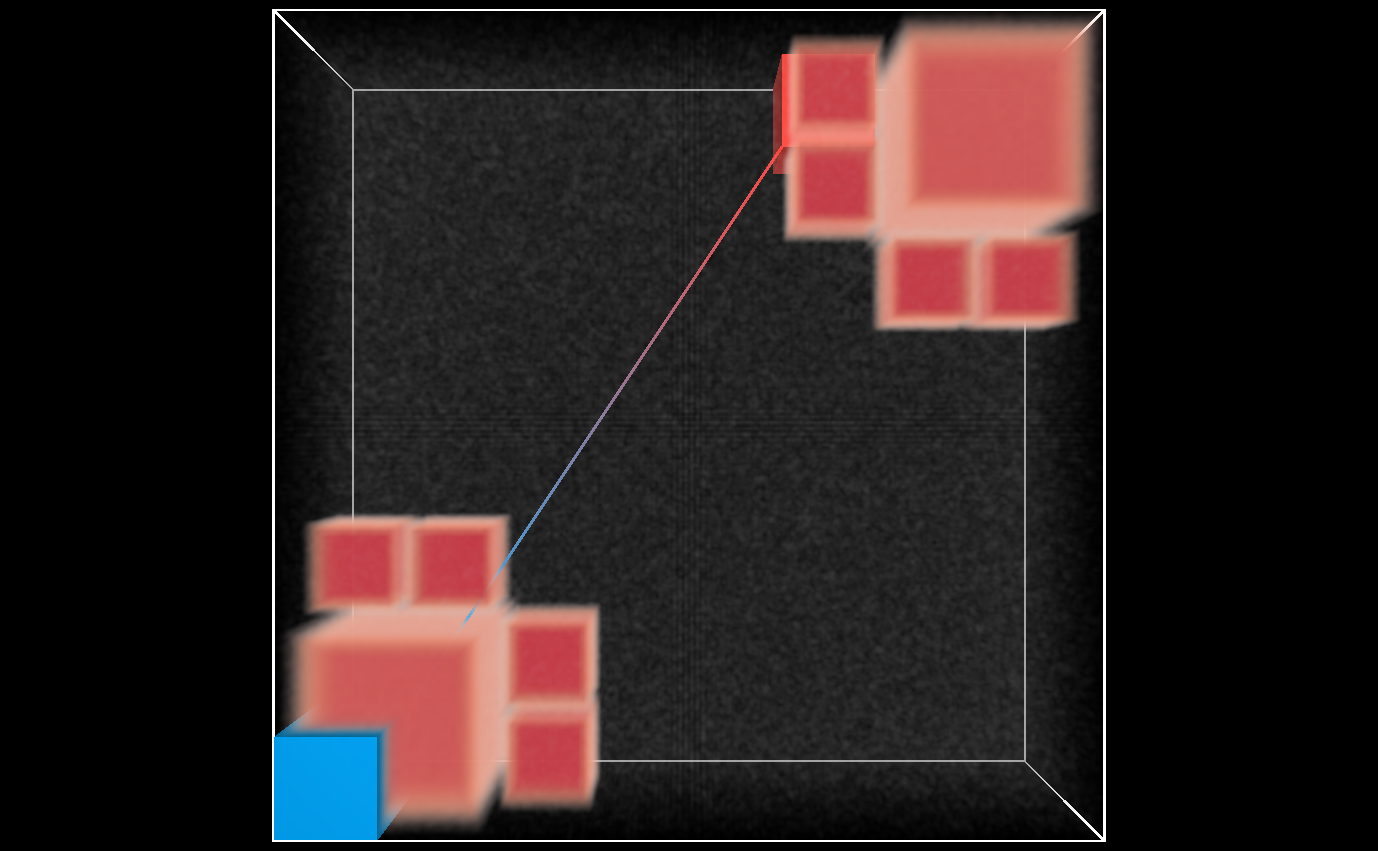}
\caption{Correlation analysis in ``Synth1'' using the PPMCC. Left: Context view (top) shows sparse brick-to-brick correlations. 3D spatial view (bottom) confirms sparsity via volume rendering the per-point correlation maxima. Red  and blue triangles indicate bricks selected for refinement. Middle: Selected bricks are shown via red and blue cubes in the spatial view (bottom). Bricks contain different amounts of points with high and low correlations, resulting in different sparsity patterns in the focus view (top). Right: Second refinement shows asymmetric pattern on top half-circle, which is due to level-wise layout of data points along the z-curve.} 
\label{fig:synthetic-4x4}
\end{figure}

\subsection{Correlation Analysis}

\subsubsection{``Synth1''} In our first experiment, we demonstrate the use of chord diagrams with BOS for an interactive analysis of the PPMCCs in ``Synth1''. Since the correlations in ``Synth1'' are known (see \cref{fig:synthetic-4x4}~bottom~left), the experiment demonstrates the potential of our approach to reveal these structures and hint towards the most prominent spatial relationships. 
``Synth1'' has been partitioned initially into $4 \times 4 \times 1$ bricks which are laid out along the context chord (see \cref{fig:synthetic-4x4}~top~left). Only few edges between a small subset of all bricks indicate high correlation, i.e., between those bricks that contain (parts of) a cluster. The user can interactively select edges to see where the bricks with low or high inter-brick correlation are located in the 3D domain. When selecting an edge in the context chord, the two bricks connected via this edge are shown in the spatial view and laid out along the half-circles of the first focus chord diagram. For the two selected bricks, the focus view shows asymmetry in the number of entities showing large correlations between the upper and the lower half-circle. This occurs because an edge has been selected in the context view that connects a region densely filled with a large cluster with a region containing two smaller clusters, i.e., correlations are between all sub-regions in the brick containing the large cluster and only a subset of sub-regions in the other brick. When going further down in the hierarchy by selecting an edge in the focus diagram, one again sees in the second focus diagram (\cref{fig:synthetic-4x4}~top~right) that both selected bricks comprise different numbers of finer bricks with high correlations. 

Notably, in all situations BOS can correctly estimate the maximum PPMCC of 1 between all bricks containing parts of the clusters. Furthermore, it can be observed that the correlations are high where the ensemble spread is low, and vice versa. This is indicated by the spread shown around the circular layout, which is due to how the synthetic data set has been created. In regions outside of the red clusters in the 3D view, ensemble values are drawn from independent and identically distributed random variables, and values are more and more linearly distributed the closer they get to a cluster center.

\begin{figure}[t]
\centering
\includegraphics[width=0.4\linewidth]{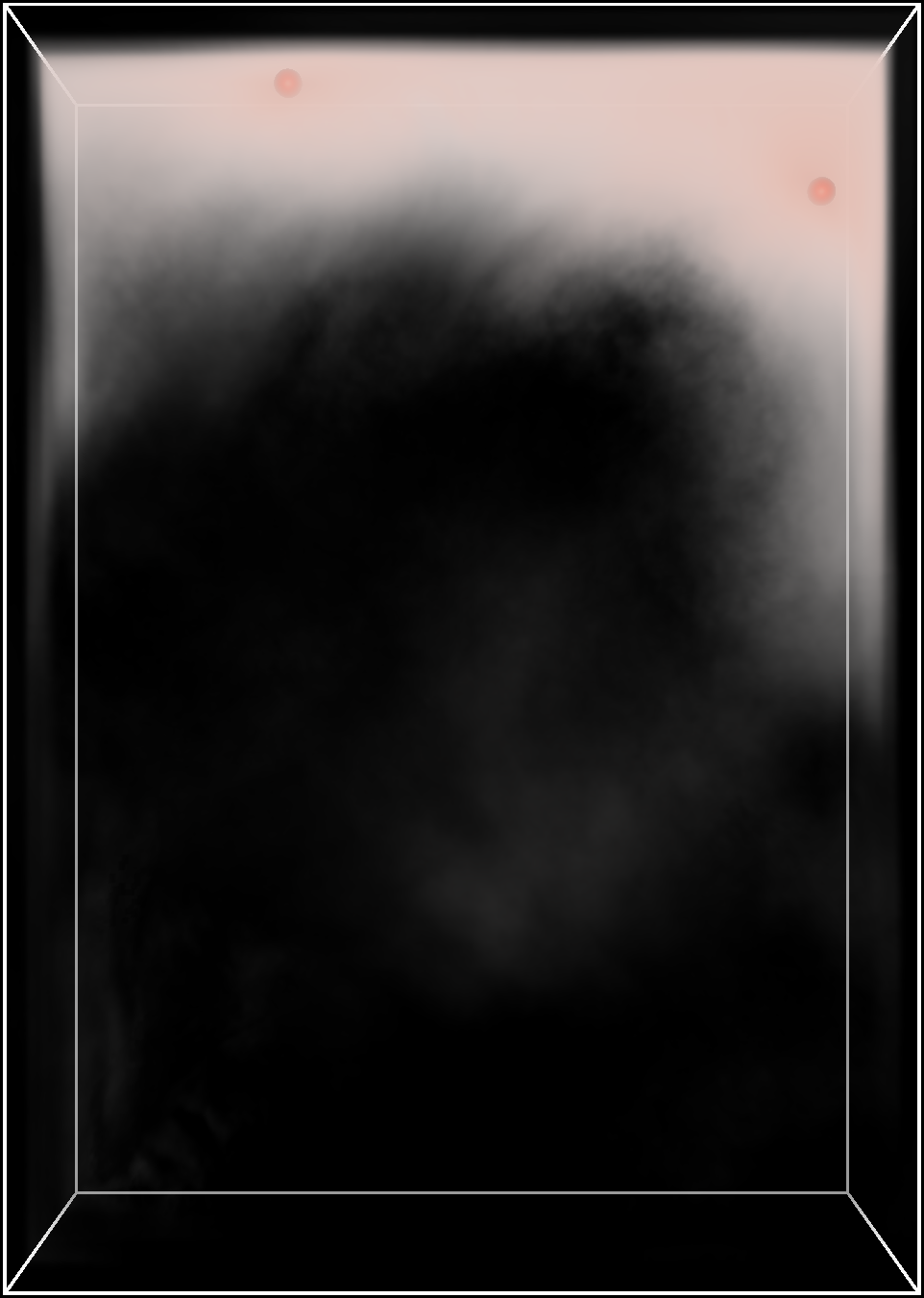} \hspace{0.03cm}
\includegraphics[width=0.4\linewidth]{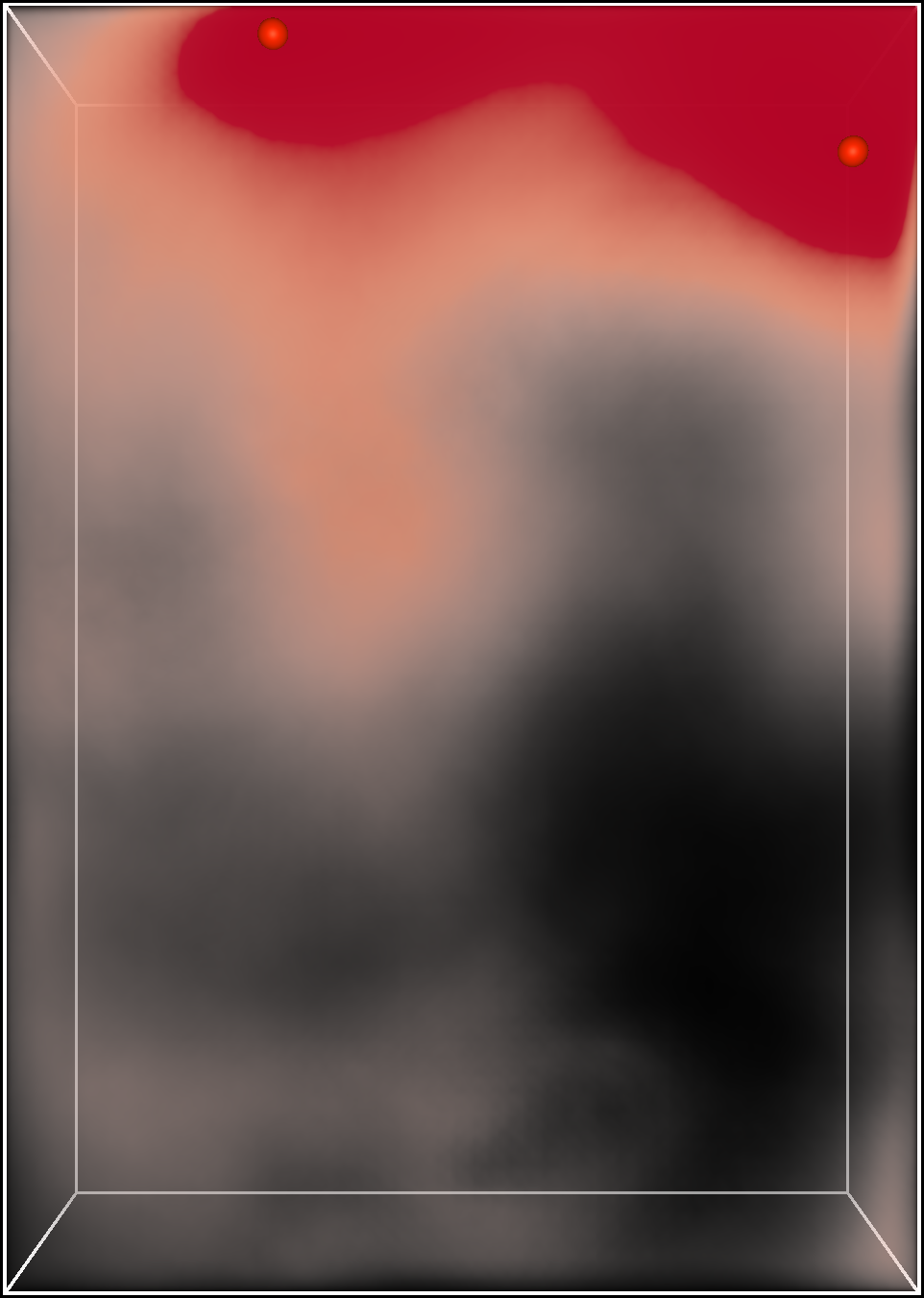} \hspace{0.05cm}
\includegraphics[width=0.12\linewidth]{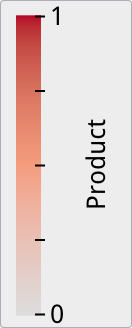}
\caption{Data set ``Necker'': At each grid point the MI in temperature \texttt{tk} between this point and the two points marked by red dots are multiplied and volume rendered. The two points correspond to the locations of the selected bricks in the context view in \cref{fig:teaser}. Left: MI in vertical layer 10. Right: MI in vertical layer 19.}
\label{fig:necker-layers}
\end{figure}

\subsubsection{``Necker''}

In a second experiment, we analyse long-range MI structures in the temperature field \texttt{tk} of ``Necker''. Therefore, we 
first suppress all correlations between bricks that are closer to each other than a selected distance threshold. A brick size of $32 \times 32 \times 20$ is used in the context view, so that all vertical levels are spanned by one brick. The focus view uses a brick size of $4 \times 4 \times 4$. This setting has been used to generate the visualizations in \cref{fig:teaser}. 

\begin{figure*}[t]
\centering
\includegraphics[height=3.1cm]{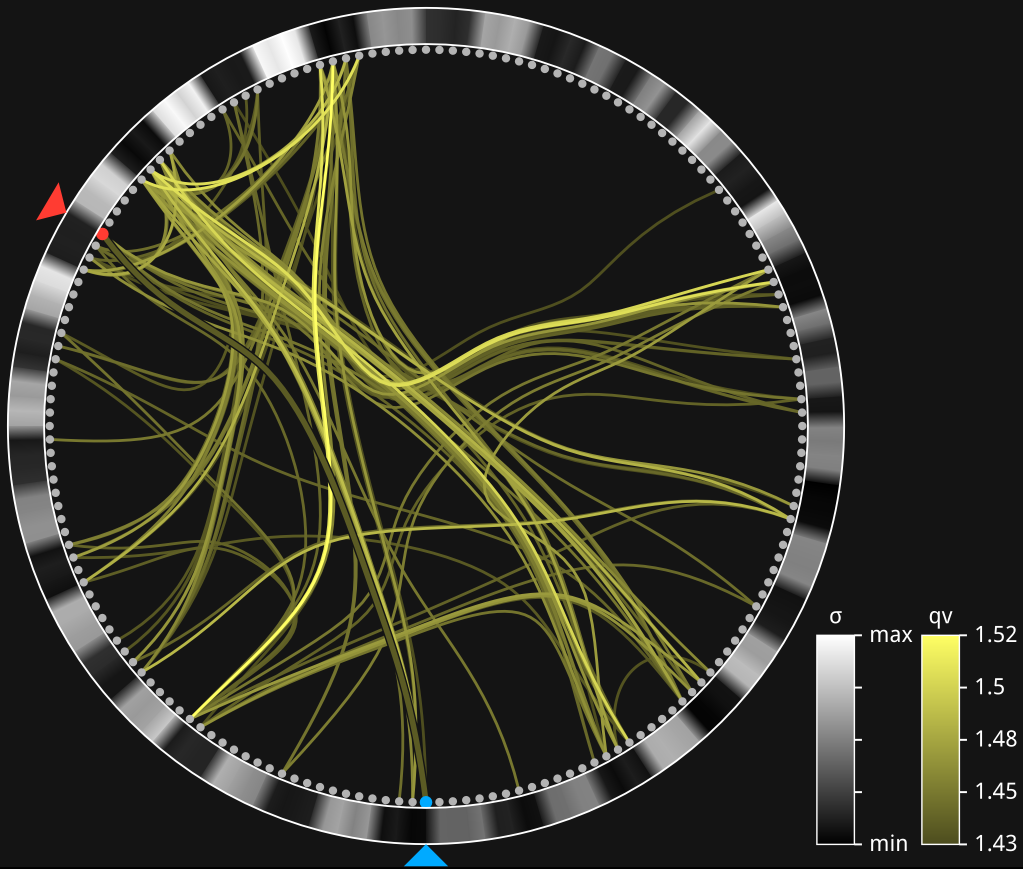}
\includegraphics[height=3.1cm]{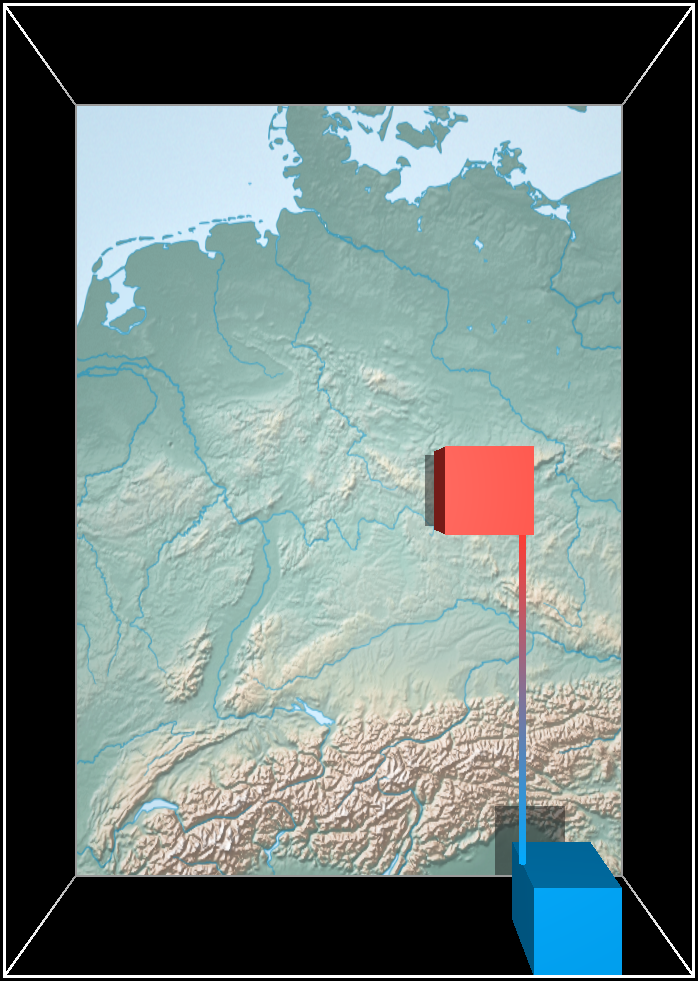}
\includegraphics[height=3.1cm]{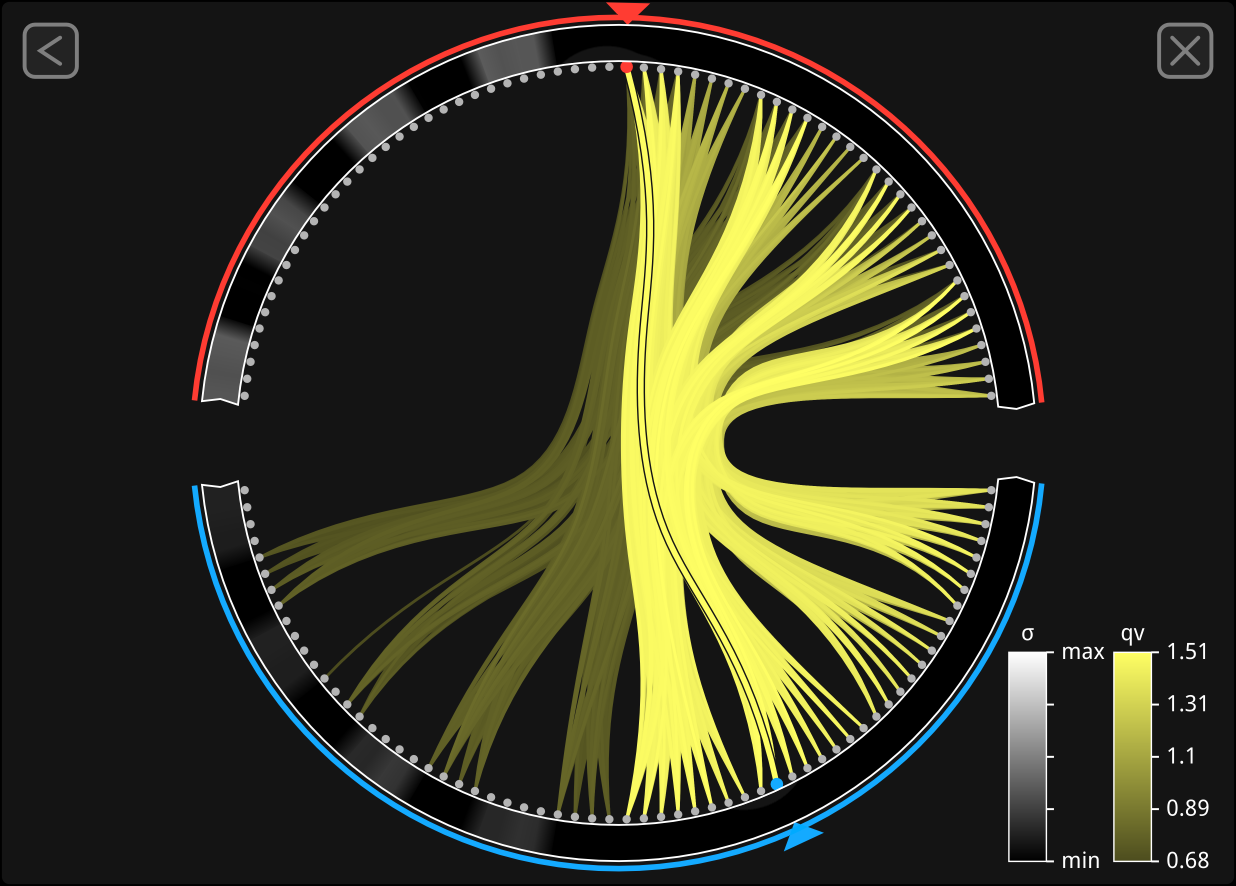}
\includegraphics[height=3.1cm]{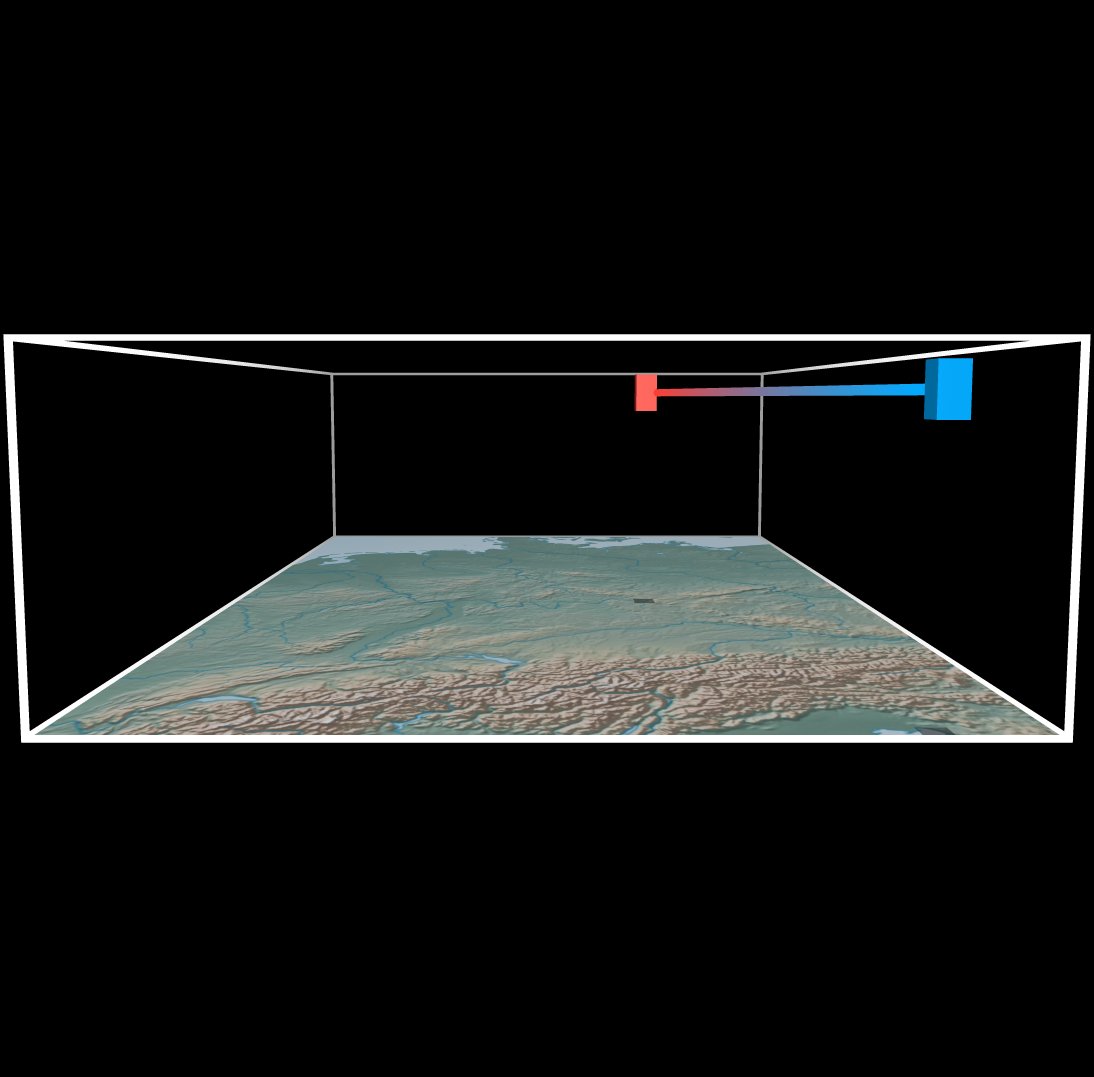}
\includegraphics[height=3.1cm]{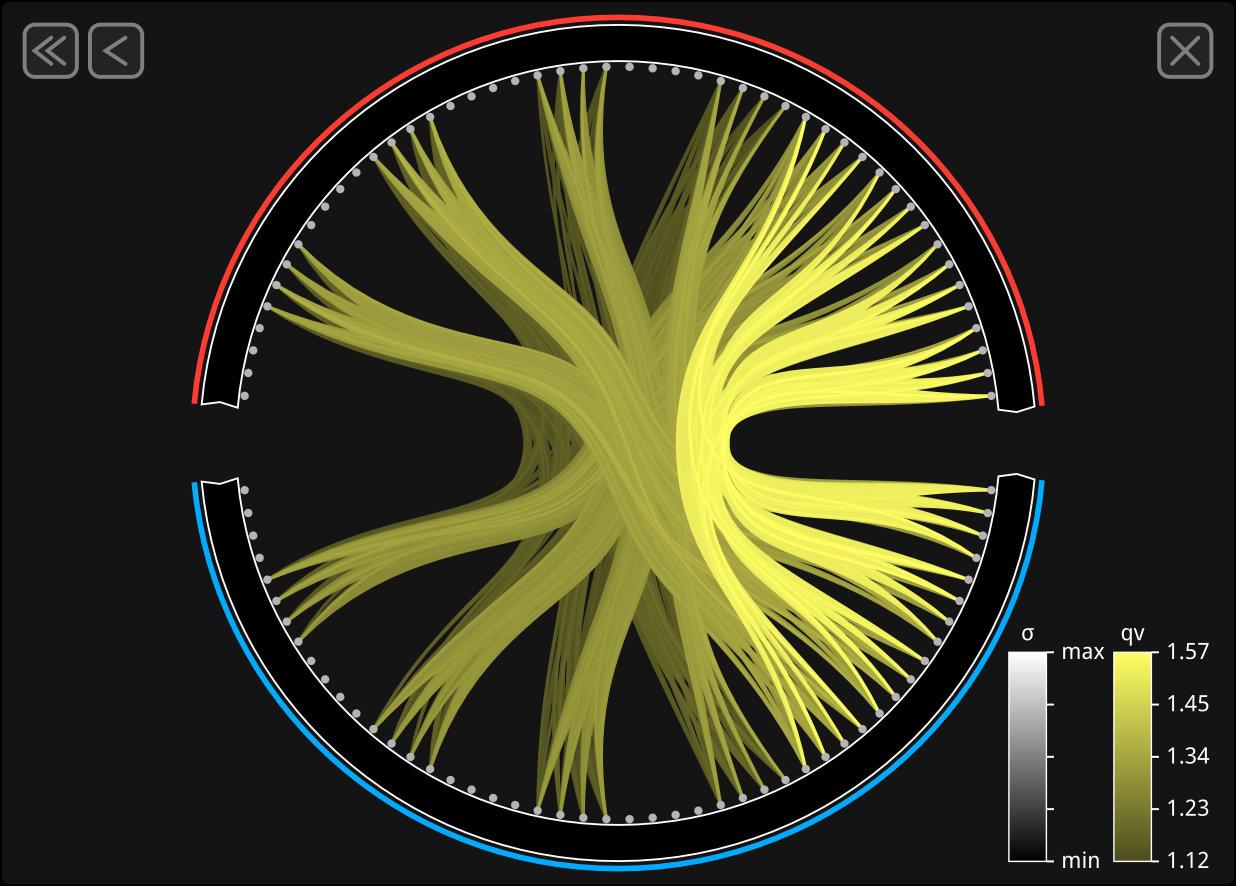}\vspace{0.08cm}
\includegraphics[height=3.1cm]{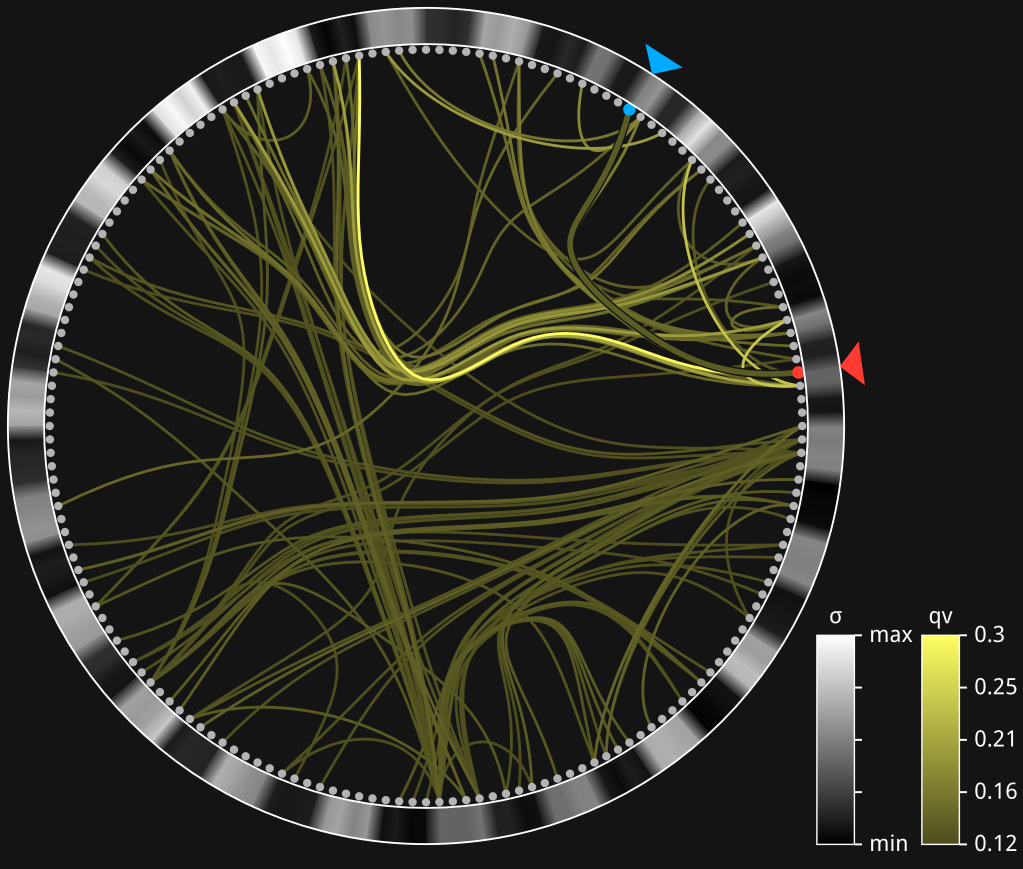}
\includegraphics[height=3.1cm]{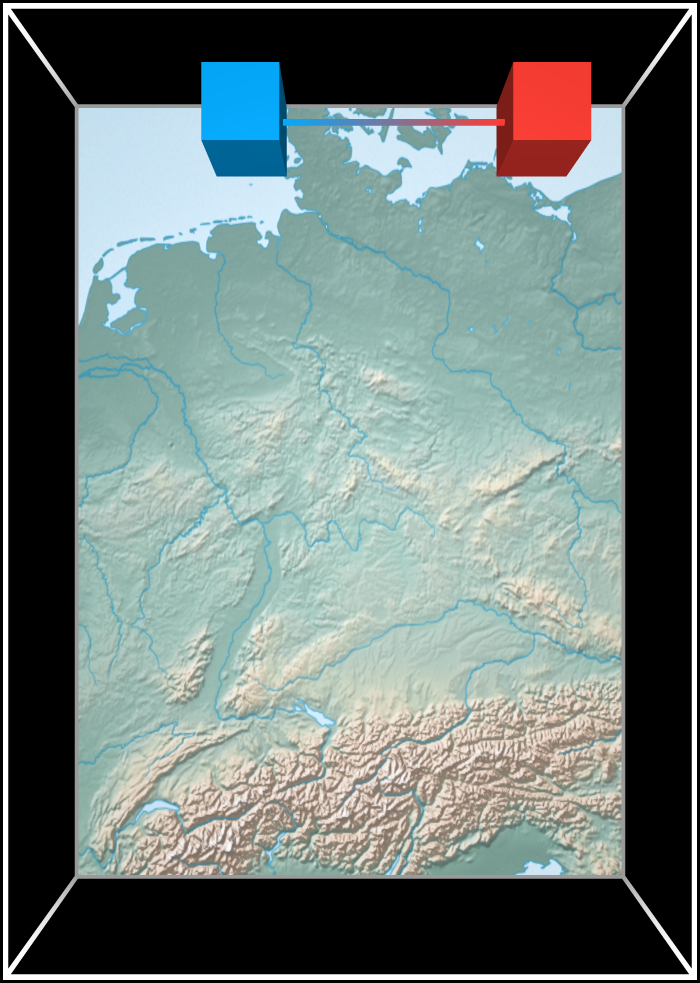}
\includegraphics[height=3.1cm]{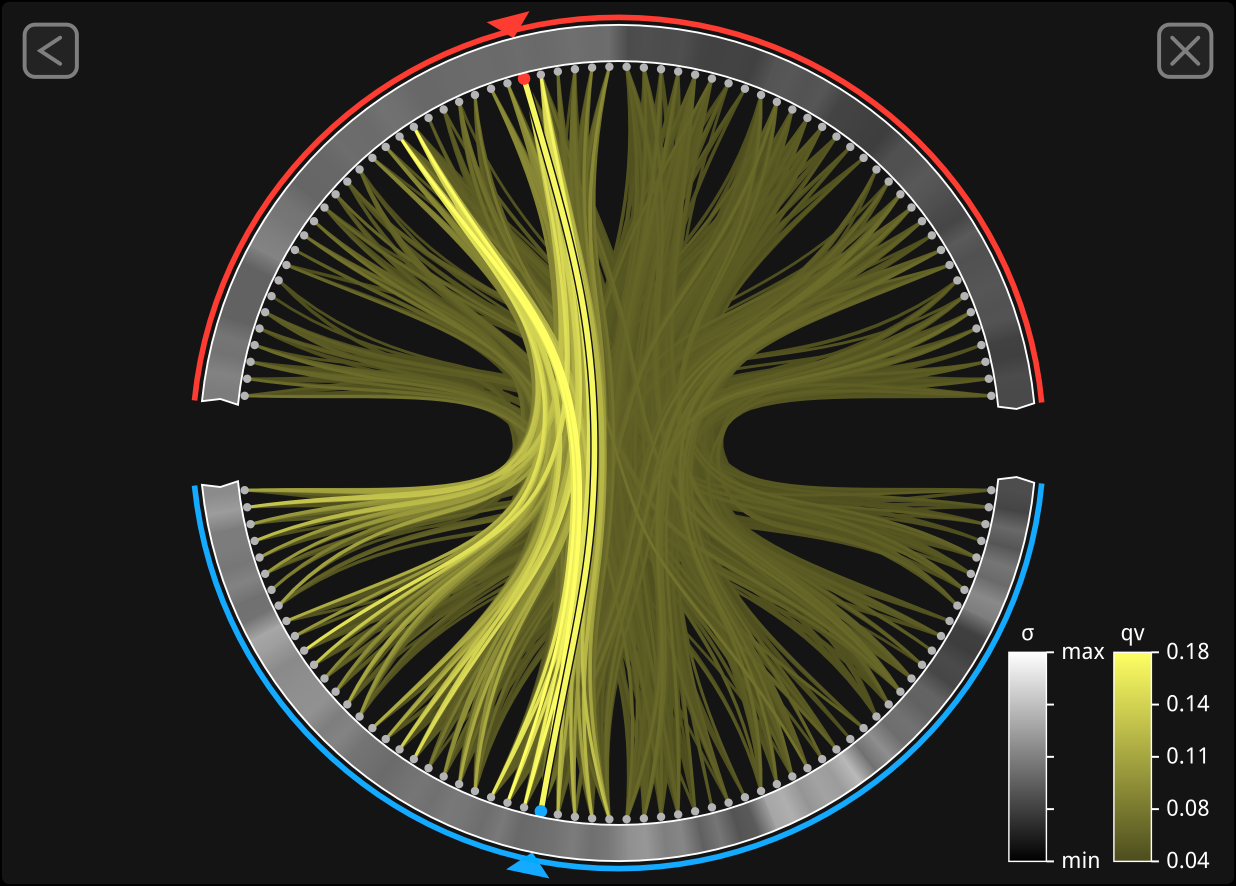}
\includegraphics[height=3.1cm]{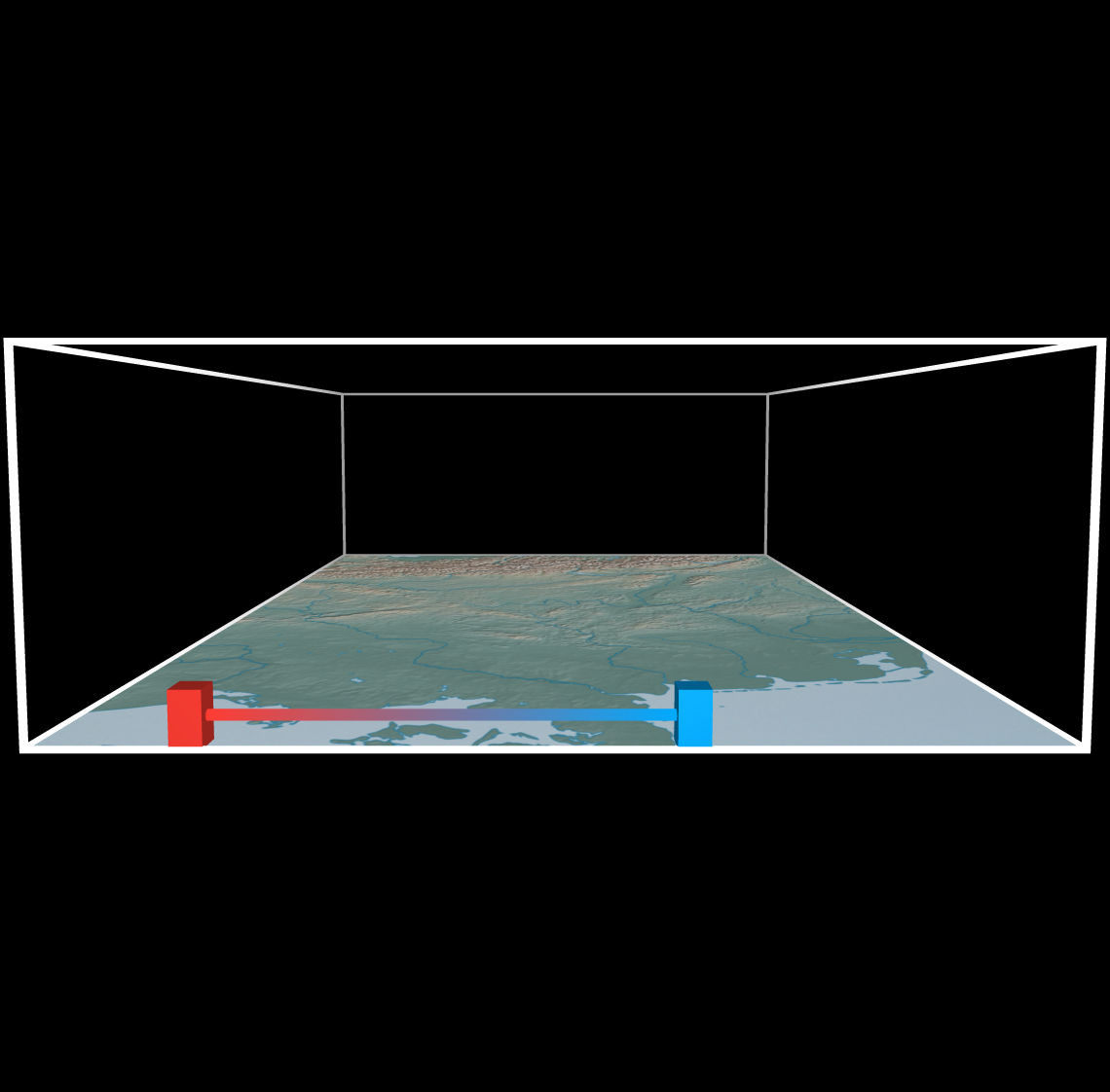}
\includegraphics[height=3.1cm]{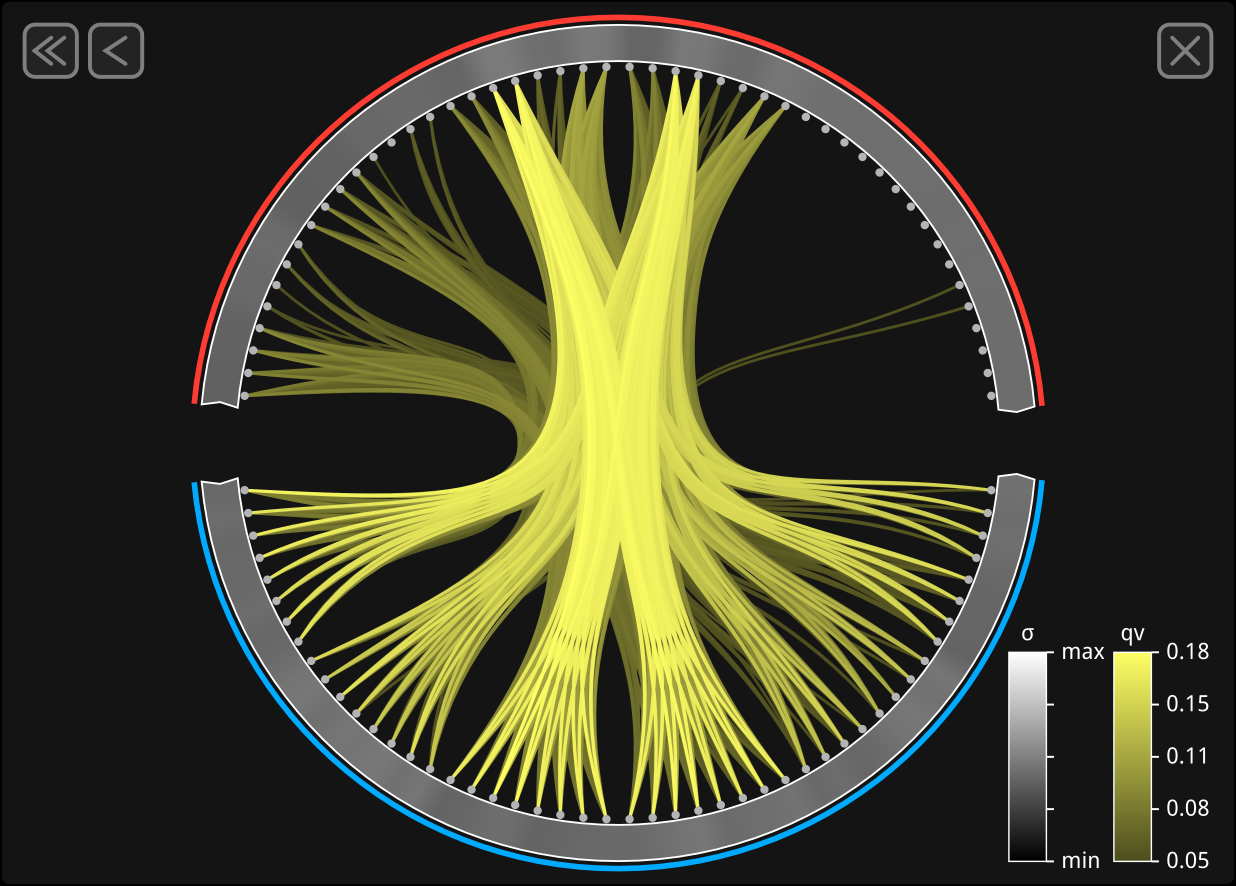}
\caption{Data set ``Necker'': Visual analysis of MI in water vapor mixing ratio \texttt{qv}.
Top, left to right: Context view 
showing correlations between bricks with distance $\ge 288$km and MI $\ge 1.43$, spatial view of selected brick pair, focus view of correlations between selected bricks, spatial view of refined bricks selected in focus view, focus view of selected refined bricks.
Bottom, left to right: Same as top, but correlations with MI $\le 0.3$ are shown. 
}
\label{fig:necker-qv}
\end{figure*}

The context view reveals multiple regions which are connected by bundles of edges indicating high correlations. One bundle that stands out is seen on the top right of the diagram. A representative edge from this cluster has been selected (indicated by the red and blue triangles), and the corresponding bricks are shown in the 3D view and further subdivided in the focus view. The selected bricks lie over the Baltic and North Sea. As can be seen in the focus view, there are 32 entity pairs with notably high correlation. By hovering over the edge with the mouse, one can see that these entities correspond to the sub-regions in the topmost vertical levels. This confirms the observation by Necker and co-workers of low correlation between levels close to the ground and the tropopause, respectively. High correlations between spatially distant layers are not present.

When interactively hovering over the edge close to the selected one, it can be seen that \texttt{tk} has higher correlations in higher vertical layers. Furthermore, a cluster of high correlation---corresponding to the inspected bundle of edges in the context view---resides in the top right corner of the 3D field. These structures are confirmed in the volume renderings of 3D MI fields in \cref{fig:necker-layers}, where MI values are computed separately between each grid point and two selected points. At each grid point, the two MI values have then been multiplied to indicate those regions where MI is high with respect to both selected points. As can be seen, higher correlations are revealed both in the most northern parts of the domain and in higher layers of the atmosphere. 
Overall, the proposed chord diagrams can effectively convey correlation structures over multiple scales in one single view. Notably, this could only be seen in correlation volumes by manual selection of many reference points.

In a second experiment, we analyze the spatial correlation structures in a different variable, the water vapor mixing ratio \texttt{qv} ($\text{kg}$ $\text{kg}^{-1}$), in \cref{fig:necker-qv}. Most interestingly, when filtering for the highest correlations in \cref{fig:necker-qv}~top, and in contrast to the findings regarding the temperature \texttt{tk}, correlations between the North and Baltic Sea are mostly absent in the context view. As can be seen, e.g., from the picked pair of bricks in the context view, spatial correlations for \texttt{qv} are higher over continental Europe. For the context view bricks spanning half of the height of the domain have been chosen. From the ensemble spread displayed in the outer ring it can be seen that the spread in lower levels is higher than in higher levels, conveyed through the striped pattern.
We have selected two bricks for closer inspection over central Germany (red) and northern Italy (blue). The highest correlations in the focus view are all confined to the highest vertical layers, with some lower correlations also to medium layers above northern Italy. This becomes visible through hovering over the high-value edges running from bottom right to top right in the focus view. At these heights, \texttt{qv} is highly correlated across large regions of continental Europe. When looking at a second focus view for a pair of selected sub-bricks (\cref{fig:necker-qv}~top right), we see high symmetry of the correlation structure between the two regions.

In \cref{fig:necker-qv}~bottom, we now filter for lower correlations (MI $\le 0.3$) to also inspect lower correlations closer to the ground. In this case, the highest correlations are now again in the region over the North and Baltic Sea.
For the bottom layers over the ocean, the correlation structures look significantly different than those over continental Europe. While they are significantly lower, the maximum is closer to the sea level. This can again be verified by hovering over high-value edges in the focus view. In the second focus view (\cref{fig:necker-qv}~bottom right), two sub-bricks close to the ground have been selected. The blue sub-brick is closer to the coast than the red one. Here, we no longer have the perfect axis symmetry from the correlations of higher levels. A few of the sub-bricks in the red region are correlated to almost all of the sub-bricks in the blue region.


\subsubsection{``Matsunobu''}

\begin{figure}[t]
\centering
\includegraphics[height=4cm]{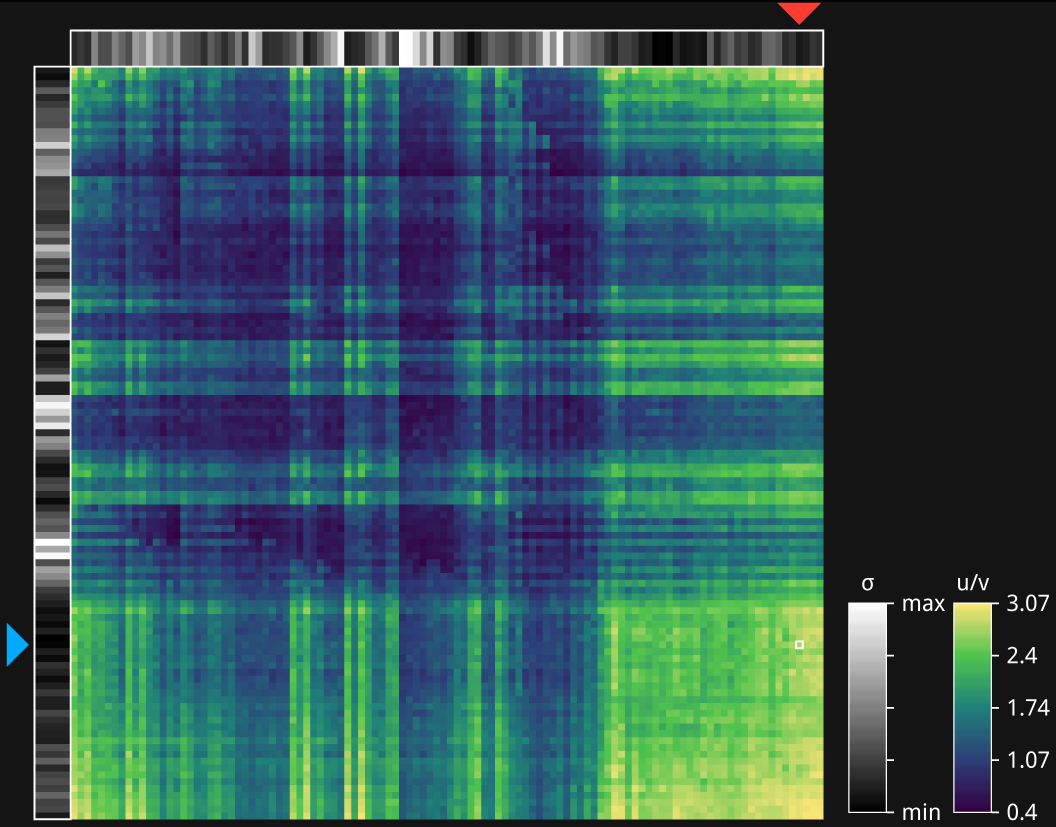}
\includegraphics[height=4cm]{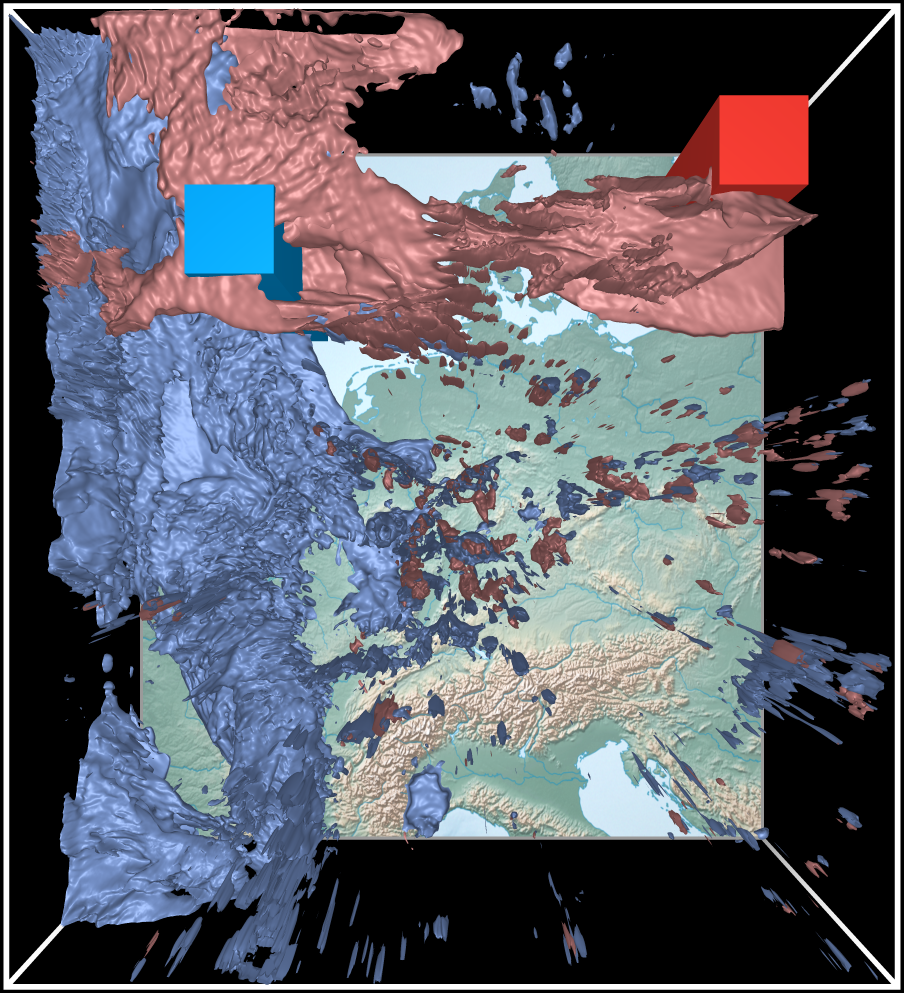}
\caption{
Left: Inter-variable correlation matrix showing MI as dependence measure between the longitudinal and latitudinal wind components \texttt{u} and \texttt{v} in ``Matsunobu''. Right: Spatial view with selected locations and isosurfaces for \texttt{u} (red) and \texttt{v} (blue) in a representative ensemble member.}
\label{fig:matsunobu-context}
\end{figure}

\begin{figure*}[t]
\centering
\begin{tikzpicture}
[align=center,node distance=0cm]
\begin{scope}[node distance=0.08cm]
  \node [inner sep=0pt] (a) {\includegraphics[height=3.1cm]{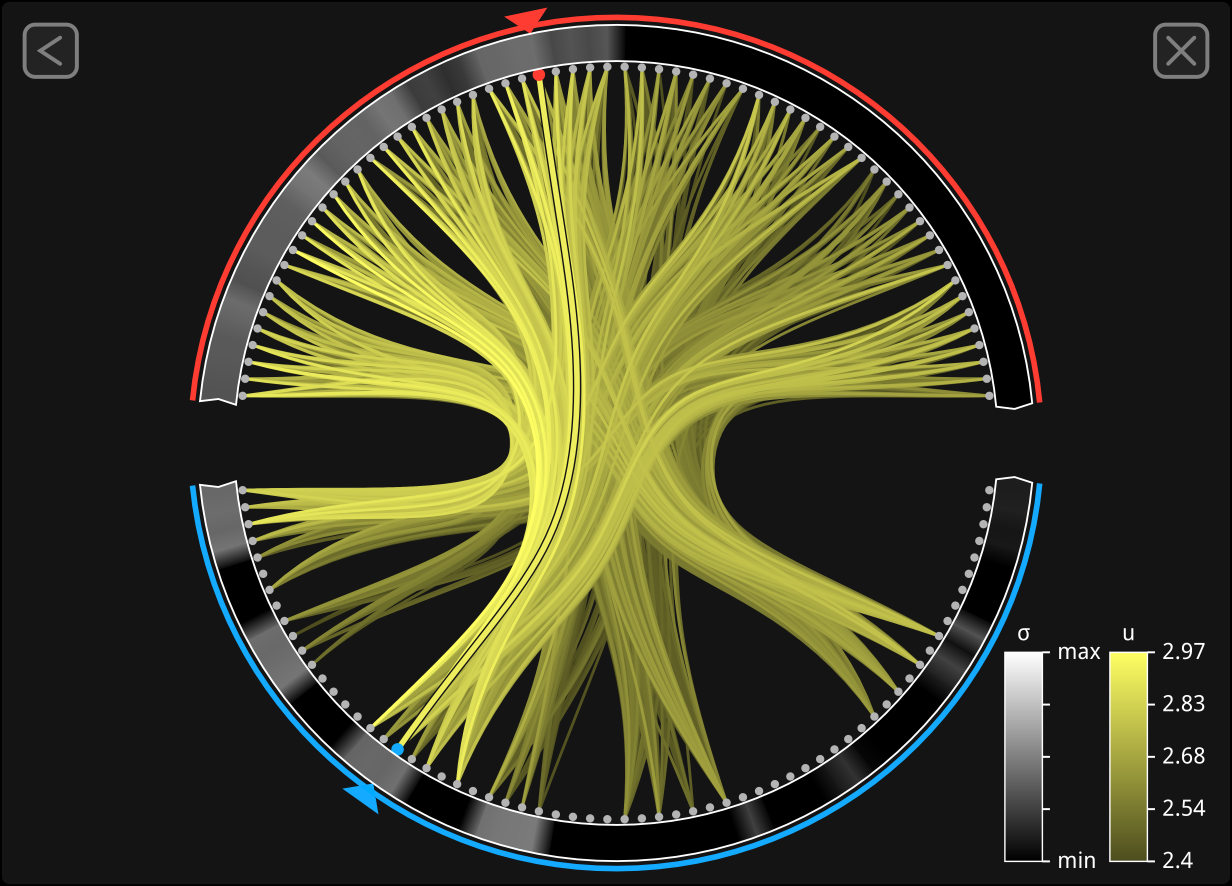}};
  \node [inner sep=0pt,right=of a] (b) {\includegraphics[height=3.1cm]{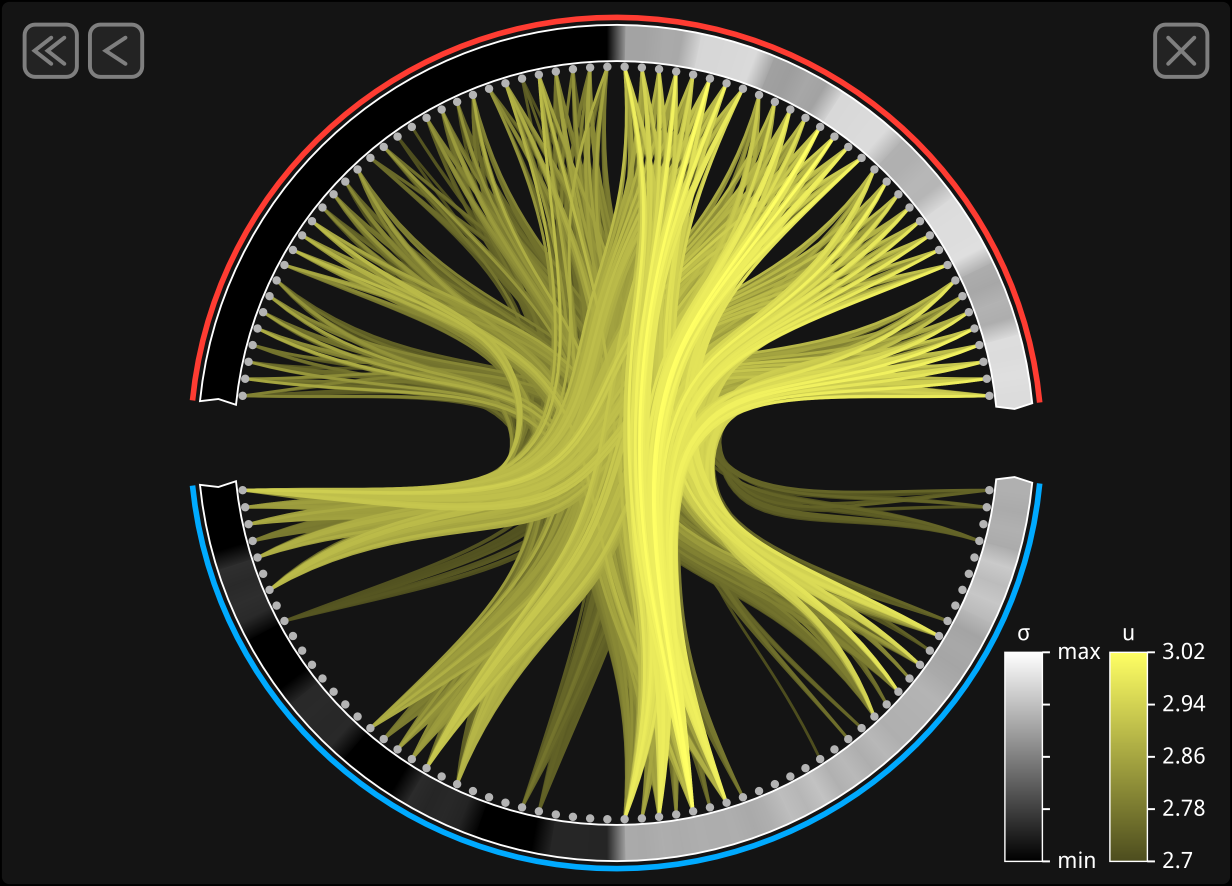}};
  \node [inner sep=0pt,below=of b] (c) {\includegraphics[height=3.1cm]{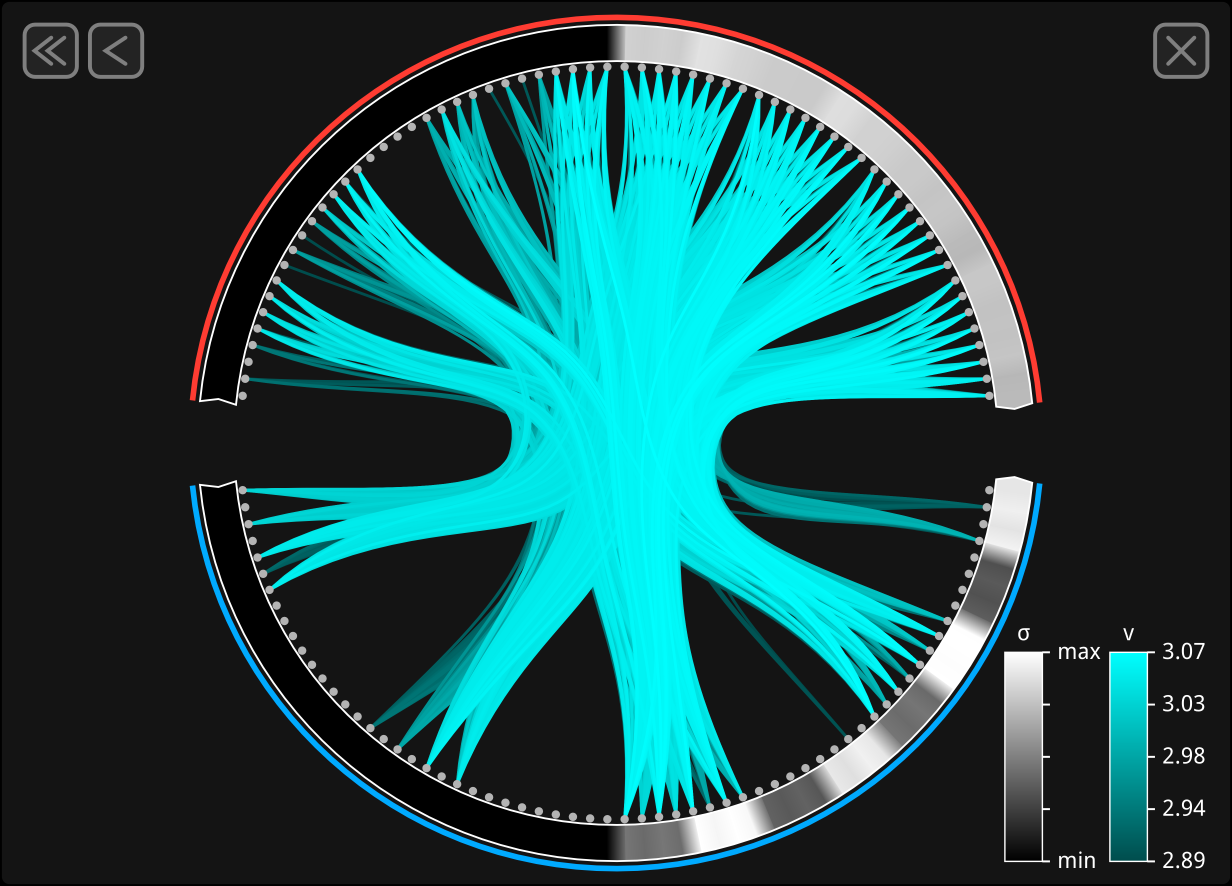}};
  \node [inner sep=0pt,left=of c] (d) {\includegraphics[height=3.1cm]{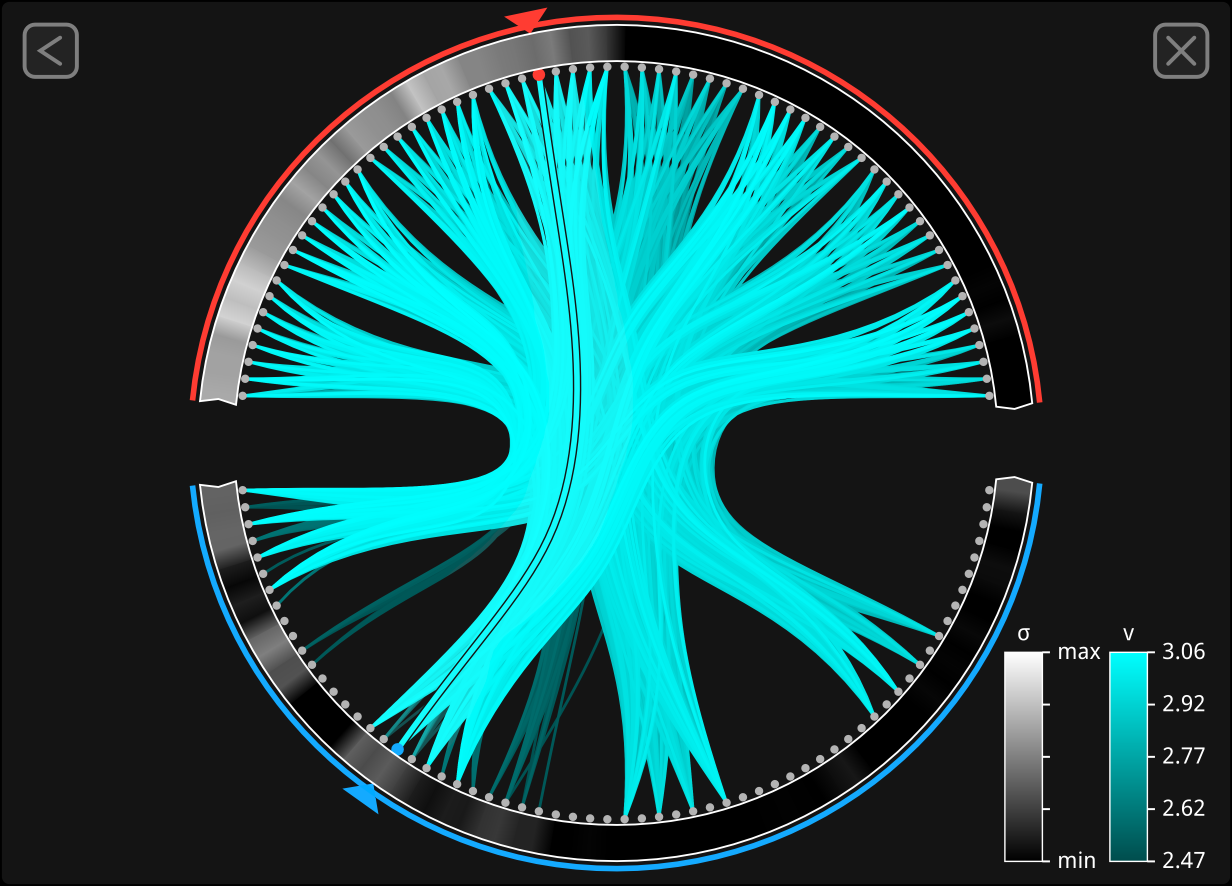}};
\end{scope}
\node[fit=(a)(b)(c)(d)](group){};
\node [left=of group,inner sep=0pt] (z) {\includegraphics[height=4.1cm]{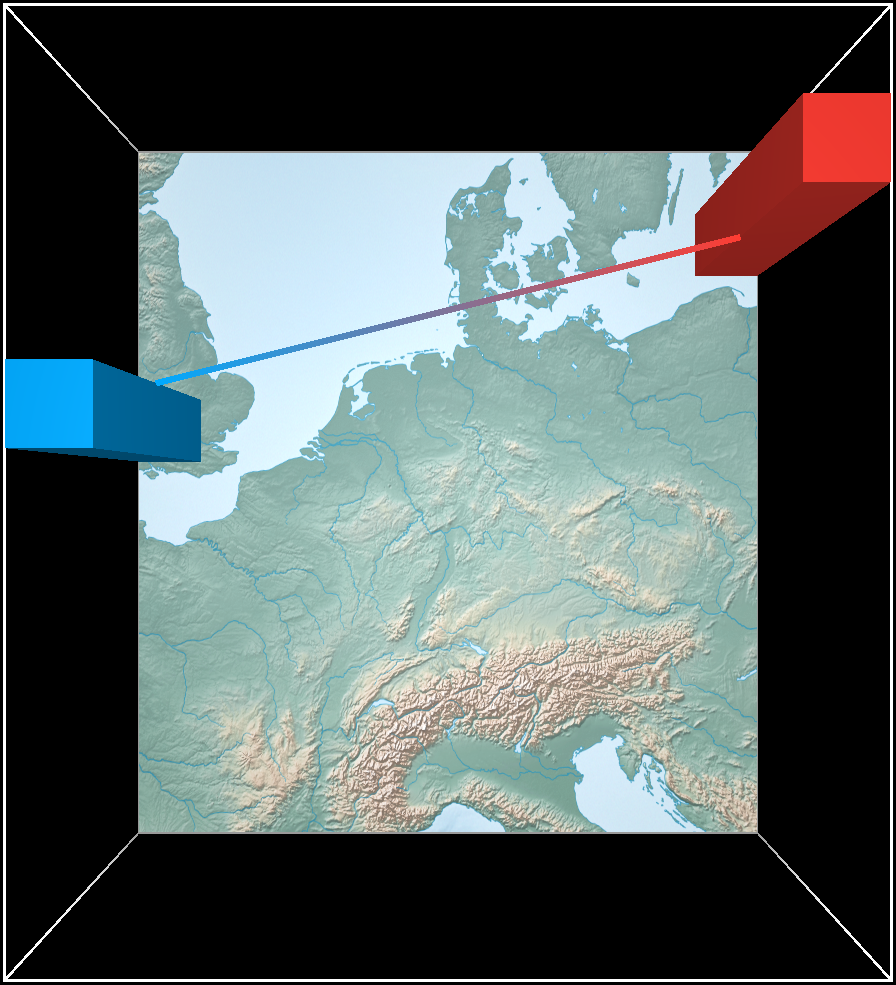}};
\node [left=0.08cm of z,inner sep=0pt] (y) {\includegraphics[height=4.1cm]{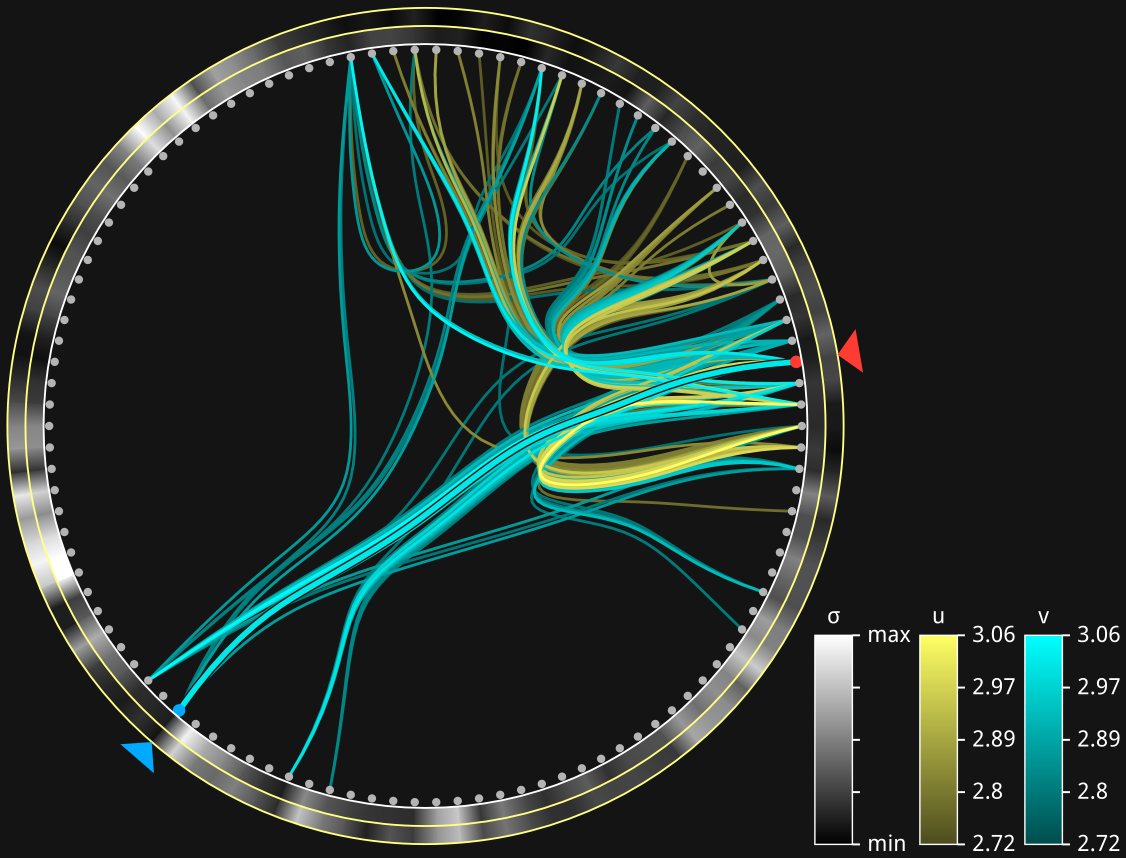}};
\end{tikzpicture}
\caption{Comparative chord visualization of MI in each of the variables \texttt{u} and \texttt{v} in ``Matsunobu''.
Left: Context diagram showing edges for brick distances $\ge 420$km.
Middle: Spatial view of selected brick pair. Right: Focus diagrams for the first and second refinement levels. 
}
\label{fig:matsunobu-comp}
\end{figure*}

In a third experiment, we first analyze the cross-correlations between the longitudinal \texttt{u} and latitudinal \texttt{v} wind components in ``Matsunobu''. \cref{fig:matsunobu-context}~left shows the corresponding inter-variable correlation matrix. 
An entry is marked with a red and blue arrow, and the selected bricks are shown in the spatial view in \cref{fig:matsunobu-context}~right. The matrix view conveys a distinct cluster with high correlation values. Bricks in this cluster are mostly located in the western and northern parts of the simulation domain, which is seen in the spatial view when hovering over the  matrix elements. Furthermore, iso-surfaces in \texttt{u} (red surface) and \texttt{v} (blue surface) corresponding to high wind speed are rendered in the spatial view.
They indicate a strong wind current in the northern and western parts. 
The standard deviations displayed next to the topmost matrix row and leftmost column show relatively low deviation between the individual ensemble members in the prominent cluster, hinting at high prediction accuracy. The high cross-correlations indicate that the individual ensemble members are strongly dependent. It is also conveyed that differences in the longitudinal and latitudinal wind components are considerably less correlated in the other regions over central Europe.  


We then shed light on the correlation structures in \texttt{u} and \texttt{v} by means of a context chord diagram (see \cref{fig:matsunobu-comp}). Again, correlations are highest in the western and northern parts over Europe. In the context view, edges for both variables are rendered jointly, sorted from front to back by their corresponding correlation value. When selecting one of the edges with the largest MI values for \texttt{v}, the corresponding focus diagrams between \texttt{u} and \texttt{v} show similar but not identical correlation structures. Overall, correlations in \texttt{v} are slightly higher and more equally distributed over the individual height layers, which is visible due to a larger amount of highly saturated edges.

\section{Limitations}


A general limitation of chord diagrams is that for more than one variable edges can occlude each other and the visualization can become cluttered. By using two distinct colors as proposed, in combination with filtering of correlations and brick pairs, this limitation can be relaxed, yet especially due to occlusions a comparative analysis remains difficult. This problem becomes even more severe when more than two variables are compared. 


BOS, on the other hand, while being able to build an accurate surrogate model with only a few samples if the underlying field is smooth, may miss correlation maxima if the field is noisy or maxima are confined to small regions. 
It is probably due to the smoothness of the correlation fields in atmospheric data that 100 BOS samples resulted in fairly accurate maxima estimations in all of our experiments. In other scenarios, however, it can be necessary to manually adapt BOS. When the user has prior knowledge on the smoothness of the sampled field, the UCB~\cite{UCB} acquisition function that is used in the Bayesian optimization process can be parameterized accordingly, i.e., new samples can be distributed to regions with high uncertainty (exploration) or regions in proximity to previous samples with high value (exploitation). The acquisition function uses a parameter $\kappa$ to control the trade-off between exploration and exploitation. In all our experiments this parameter was set to $0.5$, yet a larger value should be selected to increase exploration for fields with many local maxima and a lower value to increase the amount of exploitation and, thus, the convergence rate for smoother fields. 

It also should be mentioned that in situations where significantly more correlation samples need to be taken, the computational overhead introduced by BOS can become a limiting factor. 
When using BOS with a Gaussian process surrogate model, as in our work, the time complexity for a set of $s$ samples is $O(s^3)$. As an alternative, an approximation via sparse Gaussian processes can be used, which limits the local contribution to a set of $\hat{s}$ support points and results in a time complexity of $O(s\hat{s}^2)$ \cite{SparseGaussianProcess}. However, as can be seen in \cref{fig:sampling}~left, in our experiments we observed that beyond 100 BOS samples the accuracy of maxima estimation did not further improve considerably. Thus, 
we did not consider alternative methods for speeding up computations, yet for other types of data this can become necessary.

\section{Conclusion and Future Work}

We have proposed the use of importance-based correlation sampling to support an interactive focus+context visualization of correlation structures in large 3D ensemble fields. The maximum correlation between two sub-regions is estimated via Bayesian optimal sampling, and region-to-region maxima are visualized via a chord diagram including edge bundling to reduce visual clutter. For large ensembles, maxima are sampled from a mean-tree at a suitable resolution. Regions are linearized along a space-filling curve and mapped to the circular chord layout. The user can interactively select chord edges for a closer inspection in a focus view, and analyse correlation patterns of more than one physical variable via a chord diagram or an inter-variable correlation matrix. We have introduced optimized GPU solutions to efficiently compute prominent measures of statistical dependence in large ensembles. The use of our approach has been demonstrated with two large simulation ensembles and one synthetic data set. 


In the future, we intend to address the limitation of chord diagrams for comparing the correlations in multiple variables. Especially, we will analyse the use of set operations, i.e., set intersections, between the set of edges in the chord diagrams of different variables, to find similarities in the respective correlation fields. We believe that by using such operations the correlations in multiple fields can be compared effectively. 

The computational complexity of MI, which can hinder an interactive analysis of large ensembles, will be addressed by including neural correlation fields \cite{NeuralFieldsStatDepEnsembles}. By training a neural network to predict MI values at high speed, the sampling performance can be improved significantly. 

Furthermore, it will be interesting to include temporal correlations into chord diagrams. This, however, is challenging, since it requires to encode the time dimension and access far more data for correlation sampling. The first challenge can be addressed by using an additional circular shell around the chords in which the time axis is encoded, and in which the user can select the times that should be considered for correlation analysis. The second challenge requires the extension of the mean-tree to also cover the time domain, so that interactivity can be ensured when both spatial sub-regions and time points are selected adaptively.

Lastly, we will collaborate with domain experts to explore unknown correlation structures in large weather forecast ensembles. So far, such structures are not considered in meteorological workflows due to computational limits, even though there is agreement that they could hint at unknown relationships. Such relationships can, for instance, show dependencies between different simulation variables, or between fields that have been assimilated or simulated. Correlation structures could also be used to reveal redundancies between different ensemble members, and, thus, to reduce the number of required members. In this context, we consider in particular the analysis of the temporal evolution of correlation structures to be important.

\section*{Supplemental Materials}
\label{sec:supplemental_materials}

The code of the software implementing the visualization and interaction technique proposed in this work is made available at \url{https://github.com/chrismile/Correrender} and archived at \cite{CorrerenderZenodo}. The used synthetic correlation test data set can be generated with a script available in the GitHub repository of the application. Access to the convective-scale 1000 ensemble member simulation forecast by Necker et al.~\cite{Necker2020} used in this work can be requested from the authors of the data set. The data set by Matsunobu et al.~\cite{MatsunobuEnsemble} is publically available.




 
\bibliographystyle{IEEEtran}
\bibliography{IEEEabrv,main}

%


\section{Biography Section}
 
\vspace{11pt}

\vspace{-33pt}
\begin{IEEEbiography}[{\includegraphics[width=1in,height=1.25in,clip,keepaspectratio]{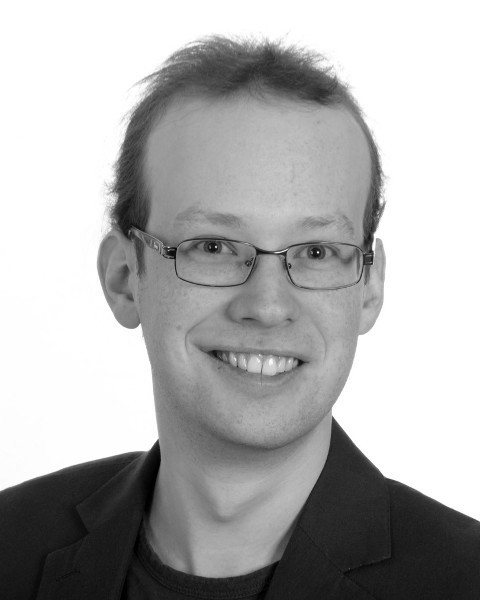}}]{Christoph Neuhauser}
is a PhD candidate at the Computer Graphics and Visualization Group at the Technical University of Munich (TUM). He received his Bachelor's and Master's degrees in computer science from TUM in 2019 and 2020. Major interests in research comprise scientific visualization and real-time rendering.
\end{IEEEbiography}

\vspace{11pt}

\vspace{-33pt}
\begin{IEEEbiography}[{\includegraphics[width=1in,height=1.25in,clip,keepaspectratio]{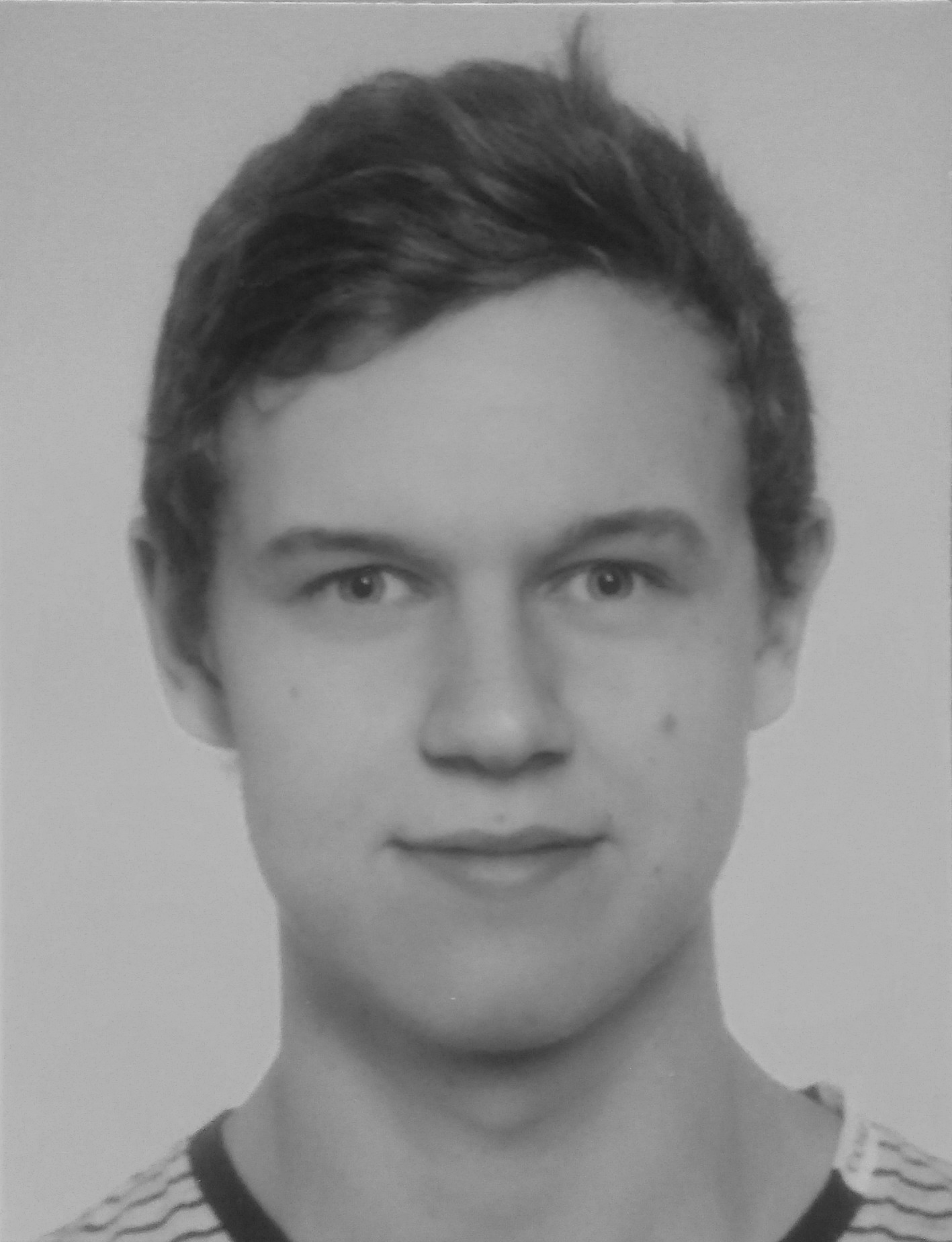}}]{Josef Stumpfegger}
is a Scientific Programmer at
the Technical University of Munich, where he obtained
his B.Sc. and M.Sc. in computer science in 2019 and 2021. His major
interests in research are scientific data visualization
for large-scale data, real-time rendering and high
performance GPU computing.
\end{IEEEbiography}

\vspace{11pt}

\vspace{-33pt}
\begin{IEEEbiography}[{\includegraphics[width=1in,height=1.25in,clip,keepaspectratio]{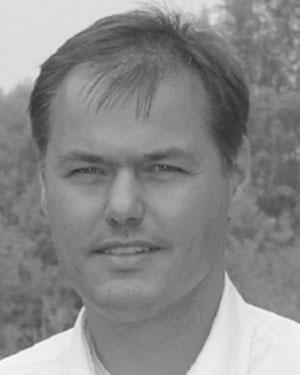}}]{R\"udiger Westermann}
studied computer science at the Technical University Darmstadt and received his Ph.D. in computer science from the University of Dortmund, both in Germany. In 2002, he was appointed the chair of Computer Graphics and Visualization at TUM. His research interests include scalable data visualization and simulation algorithms, GPU computing, real-time rendering of large data, and uncertainty visualization.
\end{IEEEbiography}

\vfill

\end{document}